%% file: localFields.tex
\documentclass[11pt,a4paper]{article}
\usepackage{amsmath,amsfonts,amssymb,amsbsy,mathtools}
\usepackage{mathrsfs,latexsym,tikz}
\usepackage{graphicx}
\usepackage{subcaption}
\usepackage{color}
\usepackage{booktabs}
\usepackage{multirow}
\usepackage{multicol}
\usepackage{alpha}
\usepackage{mathrsfs}
\usepackage{slashed}
\usepackage{arydshln}
\usepackage{bbold}

\usepackage{pdflscape}

\usepackage{ifthen}

\usepackage{macros_alpha}
\usepackage{textcomp}

\usepackage{listings}
\usepackage{add_symbols}
\usepackage{ulem}

\begin{document}
\normalem
\input{commands.tex}

\input{title.tex}

\tableofcontents

\input{intro.tex}
\input{SymEFT.tex}
\input{minimalBasis.tex}
\input{opsRenorm.tex}
\input{matching.tex}
\input{examples.tex}
\input{concl.tex}

\input{appendix.tex}

\bibliography{localFields.bib}
\bibliographystyle{JHEP}
\end{document}

%% file: commands.tex
\def\EMPTY{-EMPTY-}

\renewcommand{\L}{\mathscr{L}}
\newcommand{\Leff}{\L_\text{eff}}
\newcommand{\Llatt}{\L_\text{latt}}
\newcommand{\LeffG}{\L_\text{eff,G}}
\newcommand{\LlattG}{\L_\text{latt,G}}
\newcommand{\Lgh}{\L_\mathrm{gh}}
\newcommand{\Lgf}{\L_\mathrm{gf}}
\newcommand{\Z}{\mathcal{Z}}

\def\SQCD{S_\mathrm{QCD}}

\renewcommand{\dim}[1]{\left[#1\right]}

\newcommand{\dlatt}[2][d]{\delta #2}

\newcommand{\Jop}{J}
\newcommand{\JopE}{\Jop_\mathcal{E}}
\newcommand{\Qop}{Q}
\newcommand{\Jbase}{\mathcal{J}}
\newcommand{\JbaseE}{\Jbase_\mathcal{E}}
\newcommand{\Qbase}{\mathcal{Q}}
\newcommand{\op}{\mathcal{O}}
\newcommand{\opE}[1][]{\op_{\mathcal{E}\ifx \EMPTY#1\EMPTY \else ;#1\fi}}

\newcommand{\base}{\mathcal{B}}
\def\baseE{\base_\mathcal{E}}

\def\opEtilde{\tilde{\op}_\mathcal{E}}
\def\baseEtilde{\tilde{\base}_\mathcal{E}}

\def\lcf{J}
\newcommand{\dlcf}[1][]{J^{\ifx \EMPTY#1\EMPTY (d)\else (#1)\fi}}
\newcommand{\basedlcf}{\underline{\dlcf}}

\newcommand\AD[2][\op]{\left[\gamma_0^{#1}\right]^{(#2)}}
\newcommand\ADhat[2][\base]{\left[\hat\gamma^{#1}\right]^{(#2)}}

\newcommand{\EOMgf}{EOMs$|_\mathrm{gf}$}
\def\su#1{\ensuremath{\mathfrak{su}(#1)}}

\newcommand{\eff}[2][\mathrm{R}]{#2_{\mathrm{eff}\ifx \EMPTY#1\EMPTY \else ;#1\fi}}

\newcommand{\const}{\text{const.}}
\newcommand{\finite}{\text{finite}}

\newcommand{\DR}{\mu^{4-D}}
\newcommand{\gammaE}{\gamma_\mathrm{E}}


\newcommand{\comm}[2]{\left[#1,#2\right]}
\newcommand{\acomm}[2]{\left\lbrace #1,#2\right\rbrace}

\newcommand{\pint}[1]{\int\limits_{#1}}
\newcommand{\fint}[1]{\int\limits_0^1\D{x}}
\newcommand{\pD}{\mathcal{D}}
\newcommand{\D}{\text{d}}

\newcommand{\PI}[1][=]{\stackrel{\text{\tiny IBP}}{#1}}
\newcommand{\EOM}[1][=]{\stackrel{\text{\tiny EOM}}{#1}}
\newcommand{\BI}[1][=]{\stackrel{\text{\tiny BI}}{#1}}

\newcommand{\require}{\stackrel{!}{=}}
\newcommand{\istrue}{\stackrel{?}{=}}
\newcommand{\define}{\stackrel{\text{def}}{=}}

\def\eom{class IIa}
\def\Dslash{\slashed{D}}

\newcommand{\RGI}[2][]{#2_{\ifx \EMPTY#1\EMPTY \else #1;\fi\mathrm{RGI}}}

\newcommand{\RI}[2][]{#2_{\ifx \EMPTY#1\EMPTY \else #1;\fi\mathrm{RI}}}
\newcommand{\R}[2][]{#2_{\ifx \EMPTY#1\EMPTY \else #1;\fi\mathrm{R}}}
\renewcommand{\MS}[2][]{#2_{\ifx \EMPTY#1\EMPTY \else #1;\fi\mathrm{\overline{MS}}}}
\renewcommand{\lat}[2][]{#2_{\ifx \EMPTY#1\EMPTY \else #1;\fi\mathrm{lat}}}

\newcommand{\Sc}[2][]{#2_{\ifx \EMPTY#1\EMPTY \else #1;\fi\mathcal{S}}}

\newcommand{\bare}[2][]{#2_{\ifx \EMPTY#1\EMPTY \else #1\fi\ifx #2g 0\fi}}
\newcommand{\lattExpCont}[1]{\Exp{#1}_{\mathrm{latt}(0);\mathrm{R}}}
\newcommand{\lattExp}[2][\mathrm{R}]{\Exp{#2}_{\mathrm{latt}(a)\ifx \EMPTY#1\EMPTY \else ;#1\fi}}
\newcommand{\contExp}[2][\mathrm{R}]{\Exp{#2}_{\mathrm{cont}\ifx \EMPTY#1\EMPTY \else ;#1\fi}}
\newcommand{\latt}[2][\mathrm{R}]{#2_{\mathrm{latt}\ifx \EMPTY#1\EMPTY \else ;#1\fi}}
\newcommand{\cont}[2][\mathrm{R}]{#2_{\mathrm{cont}\ifx \EMPTY#1\EMPTY \else ;#1\fi}}

\def\lattExpRGI{\lattExp[\mathrm{RGI}]}
\def\contExpRGI{\contExp[\mathrm{RGI}]}


\def\NULL{\mathbb{0}}

\newcommand{\cev}[1]{\overset{\leftarrow}{#1}}
\renewcommand{\vec}[1]{\overset{\rightarrow}{#1}}
\newcommand{\cevvec}[1]{\overset{\longleftrightarrow}{#1}}

\def\FORM/{\texttt{FORM}}
\def\QGRAF/{\texttt{QGRAF}}
\def\MSbar{\ensuremath{\overline{\text{MS}}}%
}

\def\ctdiv{\text{cds}}

\newcommand{\UN}[2][N]{\ensuremath{\mathrm{U}(#1)\ifx \EMPTY#2\EMPTY \else _\mathrm{#2}\fi}}
\newcommand{\SUN}[2][N]{\ensuremath{\mathrm{SU}(#1)\ifx \EMPTY#2\EMPTY \else _\mathrm{#2}\fi}}
\newcommand{\Times}{\mathop{\times}}

\newcommand{\C}{\mathcal{C}}
\renewcommand{\P}{\mathcal{P}}
\newcommand{\T}{\mathcal{T}}

%% file: title.tex
\preprintno{%
\today
}

\title{\boldmath Lattice artifacts of local fermion bilinears up to $\ord(a^2)$
}


\author[ift,uam,uos]{Nikolai~Husung}
\address[ift]{Instituto de Física Teórica UAM-CSIC, C/ Nicolás Cabrera 13-15, Universidad Autónoma de Madrid, Cantoblanco 28049 Madrid, Spain}
\address[uam]{Departamento de Física Teórica, Universidad Autónoma de Madrid, Cantoblanco 28049 Madrid, Spain}
\address[uos]{Physics and Astronomy, University of Southampton,
Southampton SO17 1BJ, United Kingdom}

\vspace*{-1cm}

\begin{abstract}
Recently the asymptotic lattice spacing dependence of spectral quantities in lattice QCD has been computed to $\ord(a^2)$ using Symanzik Effective theory~\cite{Husung:2021mfl,Husung:2022kvi}.
Here, we extend these results to matrix elements and correlators of local fermion bilinears, namely the scalar, pseudo-scalar, vector, axial-vector, and tensor.
This resembles the typical current insertions for the effective Hamiltonian of electro-weak or BSM contributions, but is only a small fraction of the local fields typically considered.
We again restrict considerations to lattice QCD actions with Wilson or Ginsparg-Wilson quarks and thus lattice formulations of QCD without flavour-changing interactions realising at least $\mathrm{SU}(\Nf)_\mathrm{V}\times\mathrm{SU}(\Nb|\Nb)_\mathrm{V}$ flavour symmetries for $\Nf$ sea-quarks and $\Nb$ quenched valence-quarks respectively in the massless limit.
Overall we find only few cases $\hat{\Gamma}$, which worsen the asymptotic lattice spacing dependence $a^n[2b_0\gbar^2(1/a)]^{\hat{\Gamma}}$ compared to the classically expected $a^n$-scaling.
Other than for trivial flavour quantum numbers, only the axial-vector and much milder the tensor may cause some problems at $\ord(a)$, strongly suggesting to use at least tree-level Symanzik improvement of those local fields.
\end{abstract}


\maketitle

%% file: intro.tex
\section{Introduction}
A major systematic uncertainty of lattice QCD predictions arises from the continuum extrapolation $a\searrow 0$.
In previous publications~\cite{Husung:2021mfl,Husung:2022kvi} we have discussed the impact of quantum corrections on the $\ord(a^{\nmin})$ lattice artifacts of spectral quantities in lattice QCD, when using Wilson~\cite{Wilson:1974,Wilson:1975id} or Ginsparg-Wilson~\cite{Ginsparg:1981bj} (GW) quarks.
Here, $a$ denotes the small but non-zero lattice spacing and $\nmin$ is a positive integer that depends on the chosen lattice discretisation (nowadays one typically finds $\nmin=2$).
In an asymptotically-free theory like QCD the asymptotic lattice-spacing dependence takes the form\footnote{Actually further factors of $\log(2b_0\gbar^2(1/a))$ may arise multiplying the overall power law assumed here.
For the contributions from the action of Wilson or GW quarks we found such issues only for mixed actions or quenched QCD~\cite{Husung:2022kvi} in the range of numbers of flavours explored.} $a^{\nmin}[2b_0\gbar^2(1/a)]^{\hat{\Gamma}_i}$ rather than a simple integer power-law $a^{\nmin}$, that one would expect for a classical field theory.
Here, $\gbar(1/a)$ is the running coupling at renormalisation scale $\mu=1/a$ and $\hat{\Gamma}_i$ is related to the 1-loop anomalous dimensions of higher-dimensional operators, which describe the asymptotically-leading lattice-spacing dependence using Symanzik's Effective theory~\cite{Symanzik:1979ph,Symanzik:1981hc,Symanzik:1983dc,Symanzik:1983gh} (SymEFT).

In this paper we will focus on the additional powers $\hat{\Gamma}_i^{\Jop}$ arising for local fields.
This excludes integrated correlation functions such as moments~\cite{Bochkarev:1995ai,HPQCD:2008kxl} or QCD contributions to muon $g-2$, see e.g.~\cite{Aubin:2022hgm,FermilabLattice:2022izv,RBC:2023pvn,Kuberski:2024bcj}.
The strategy outlined here should be applicable to any local field of interest.
Each local field involved in a $n$-point function introduces its own set of operators causing additional powers $\hat{\Gamma}_i^{\Jop}$ associated to those operators to become relevant.
We will focus here on fermion bilinears $J$ of mass-dimension~3, namely the scalar, pseudo-scalar, vector, axial-vector, and tensor.
For those cases, the mass-dimension~4 operator basis relevant for $\ord(a)$ corrections has been discussed earlier~\cite{Luscher:1996sc,Capitani:1999ay,Capitani:2000xi,Bhattacharya:2003nd,Bhattacharya:2005rb} with explicit improvement in mind and not its impact on scaling due to the powers $\hat{\Gamma}_i^{\Jop}$.
Here we aim at precisely those powers for both $\ord(a)$ as well as $\ord(a^2)$ and extend the applicability of our original results to matrix elements and correlators of the local fields considered and thus beyond spectral quantities.
Observables that become accessible to this kind of analysis are, among others, decay constants, and form-factors for \mbox{(semi-)}leptonic decays of QCD eigenstates.
See the FLAG reviews~\cite{FlavourLatticeAveragingGroupFLAG:2021npn} for the status of lattice QCD results for these and other quantities.
Apart from the scalar, we consider both flavour-singlets and non-singlets.
Local operators with vacuum quantum numbers require additive renormalisation from the identity operator, cf.~\fig{fig:P}, multiplied by an appropriate power in the quark masses to get the canonical mass-dimensions right.
While this does not pose a large problem for the bilinear itself or its higher-dimensional counterparts, the proper treatment of contact terms with operators from the basis of the SymEFT action becomes very tedious.
To avoid these complications as well as the need for more general formulae in \sect{sec:OpsRenorm} we will consider the scalar only with non-trivial flavour quantum numbers.

Most of our notation has been introduced in~\cite{Husung:2019ytz,Husung:2021mfl,Husung:2022kvi} and we will not go into too much detail here.
Nonetheless, to make the differences between the new contributions and those originating from the lattice action more apparent, we will first provide a brief recap of the SymEFT approach in \sect{sec:SymEFT}, before introducing in \sect{sec:minBasis} the minimal on-shell bases needed for each local field.
From the 1-loop renormalisation of this operator bases in \sect{sec:OpsRenorm}, we then derive the lower bounds on the additional powers $\hat{\Gamma}_i^{\Jop}\geq \hat{\gamma}_i^{\Jop}$ introduced by the lattice artifacts of the various local fields, where $\hat{\gamma}_i^{\Jop}$ can be extracted from the corresponding 1-loop anomalous-dimension matrix.
We also show in \sect{sec:TLmatching} how to acquire the (tree-level) matching coefficients to take any overall suppression by (at least) one power of $\gbar^2(1/a)$ into account.
The use cases of the results obtained here are presented in \sect{sec:examples} for two simple examples in Wilson QCD.
A detailed discussion of the results and their applicability takes place in \sects{sec:discussion} and \ref{sec:limitations} respectively.
The general outcome is then summarised in \sect
{sec:conclusion}.

%% file: SymEFT.tex
\section{Symanzik Effective theory for local fields}\label{sec:SymEFT}
The SymEFT describes lattice artifacts as a perturbation around the continuum fields, i.e., for the Lagrangian of the Effective Field Theory we may formally write
\begin{equation}
\Leff=\L+a^{\nmin}\sum_i\omega_i^{\Qop}(\bare{g}^2)\Qop_i^{(\nmin)}+\ord(a^{\nmin+1}),\quad [\Qop_i^{(\nmin)}]=4+\nmin,\label{eq:Leff}
\end{equation}
where $\L$ is the continuum Lagrangian of Euclidean QCD
\begin{align}
\L&=-\frac{1}{2g_0^2}\tr(F_{\mu\nu}F_{\mu\nu})+\bar{\Psi}\left(\gamma_\mu D_\mu(A)+M\right)\Psi+\bar{\Phi}\left(\gamma_\mu D_\mu(A)+M\right)\Phi,\\
\Psi&=(\psi_1,\ldots,\psi_{\Nf})^T,\quad \Phi=(\psi_{\Nf+1},\ldots,\psi_{\Nf+\Nb},\phi_1,\ldots,\phi_{\Nb})^T\nonumber
\end{align}
with $\Nf$ sea quarks, $\Nb$ valence quarks, covariant derivative $D_\mu(A)=\partial_\mu+A_\mu$, $A_\mu\in\mathrm{su}(N)$ and field-strength tensor $F_{\mu\nu}=[D_\mu,D_\nu]$.
Notice that we choose to implement the valence quarks explicitly as \emph{additional} quenched flavours, i.e., $\phi$ are the commuting ghosts formally introduced to cancel any contributions of the valence quarks to the sea.
This leaves the freedom of choosing different discretisations for both sea and valence quarks even for flavours that are already present in the sea, see \cite{Husung:2022kvi} for a discussion of mixed actions in SymEFT.
Here and in the following, the superscript~${}^{(d)}$ always denotes the increase $d$ of the canonical mass-dimension of the operator describing the lattice artifacts compared to the continuum field of interest, or simply, the order in the lattice spacing at which those corrections become relevant.
In contrast to our earlier description for spectral quantities, we use here (explicitly) the enlarged minimal operator basis $\Qop^{(d)}=\op^{(d)}\cup\opE^{(d)}$, where $\op^{(d)}$ is the minimal on-shell operator basis and $\opE^{(d)}$ is a minimal basis of operators that vanish by the \emph{classical} equations of motion (EOM)
\begin{align}
\bar{\Psi}\cev{D}_\mu\gamma_\mu&=\bar{\Psi}M,\quad \gamma_\mu D_\mu\Psi=-M\Psi,\quad \bar{\Phi}\cev{D}_\mu\gamma_\mu=\bar{\Phi}M,\quad \gamma_\mu D_\mu\Phi=-M\Phi,\nonumber\\
D_\mu F_{\mu\nu}^a&=g_0^2\bar{\Psi}\gamma_\nu T^a\Psi+g_0^2\bar{\Phi}\gamma_\nu T^a\Phi.
\end{align}
In accordance with the notation in \cite{Kluberg-Stern:1975ebk}, we will from here on refer to such operators as \emph{\eom{}}.
The set of operators $\Qop^{(d)}$ complies with the symmetries of the lattice action and $\omega_i^{\Qop}$ are the corresponding (bare) matching coefficients.
A listing of the minimal on-shell basis $\op^{(d)}$ for Wilson or GW quarks up to mass-dimension 6 can be found in the appendix~\ref{sec:onshellActionBasis}.
Use of the enlarged operator basis is necessary due to the presence of contact terms after the expansion of the SymEFT in the lattice spacing $a$, when operators of the SymEFT action and local fields coincide in spacetime.
In section~\ref{sec:OpsRenorm} we will explain in detail the full strategy used here to achieve renormalisation to 1-loop order and how to deal with those \eom{} operators.

As for the lattice action, the discretisation of each local field $\Jop$ introduces additional lattice artifacts.
In the SymEFT this can be expressed as
\begin{equation}
\Jop_\mathrm{eff}(x)=\Jop(x)+a^{\nmin}\sum_i\nu_i^{\Jop}(\bare{g}^2)\Jop^{(\nmin)}_i(x)+\ord(a^{\nmin+1}),\quad [\Jop^{(d)}_i]=[\Jop]+d,\label{eq:Jeff}
\end{equation}
where again $\Jop^{(d)}_i$ form a minimal on-shell basis constrained by the transformation properties of the local field in the lattice theory with (bare) matching coefficients $\nu_i^{\Jop}$.
In principle, $\Jop_\mathrm{eff}$ also contains higher-dimensional \eom{} operators, which we neglect as they will only be relevant here to subleading order in the lattice spacing.
Notice, that $\nmin$ is not necessarily the same for the Lagrangian and some local field due to potentially more or less restrictive symmetry constraints.
For simplicity we will assume that both are the same (or otherwise choose the lower value).

\def\nullvec{\mathbf{0}}
\paragraph{Example: Pion decay constant}
Let us assume that we want to predict the asymptotic lattice-spacing dependence of the decay constant of a pion at rest via
\begin{align}
Z_{\hat{\mathrm{A}}}(g_0,a\mu)\langle 0|\hat{A}_0^{ud}(0)|\pi(\nullvec)\rangle=[m_\pi f_\pi](a),
\end{align}
where the superscripts indicate the flavour content of the local field in anticipation of the notation used in section~\ref{sec:minBasis}, $Z_{\hat{\mathrm{A}}}$ is the renormalisation factor of the axial-vector current, which will depend on the particular discretisation chosen as well as the scheme, and $|\pi(\nullvec)\rangle$ is an asymptotic pion at rest.
Ignoring all other systematics, $[m_\pi f_\pi](a)$ should only get contributions of lattice artifacts from the action and the axial-vector current.
This assumes that the lattice artifacts belonging to the pion-interpolating operator trivially cancel out when creating the asymptotic pion state.

(Formally) expanding the counterpart in our SymEFT around the small lattice spacing,\footnote{This is just the usual strategy like in any other effective field theory, where the ``new-physics'' cut-off is just an external parameter.
Making such an expansion in the lattice theory would be wrong because the integrations over the Brillouin zone and the asymptotic expansion in the lattice spacing --- acting as the UV regulator of the theory --- do not commute.} we arrive at an expression in continuum QCD
\begin{align}
\frac{[m_\pi f_\pi](a)}{\lim\limits_{a'\searrow 0}[m_\pi f_\pi](a')}&=1+a^{\nmin}\sum_i d_i^{\hat{\mathrm{A}}}(a\mu,\gbar(\mu))\frac{\big\langle 0\big| \left(\mathrm{A}_0^{ud}\right)_{i;\overline{\text{MS}}}^{(\nmin)}(0)\big|\pi(\nullvec)\big\rangle}{m_\pi f_\pi}\label{eq:fpiExample}\\
-a^{\nmin}\sum_i&c_i^\Qop(a\mu,\gbar(\mu))\int\rmd^Dz\,\frac{\langle 0|(\mathrm{A}_0^{ud})_{\MSbar}(0)\MS[i]\Qop^{(\nmin)}(z)|\pi(\nullvec)\rangle_\mathrm{c}}{m_\pi f_\pi}+\ord(a^{\nmin+1}),\nonumber
\end{align}
where $d_i^{\hat{\mathrm{A}}}$ and $c_i^\Qop$ are the renormalised matching coefficients directly related to the bare ones in \eqs{eq:Leff} and \eqref{eq:Jeff}, and the subscript $\langle\ldots\rangle_\mathrm{c}$ indicates that only connected pieces of $\Qop_i$ contribute.
Beyond tree-level, the matching coefficients obviously depend on the chosen renormalisation scheme and the renormalisation scale at which the matching took place.
Throughout this paper we assume $\mu=1/a$, which is the relevant scale of lattice artifacts at which the SymEFT is matched to the lattice action.
Of course, the $d_i^{\hat{\mathrm{A}}}$ will depend on the particular discretisation chosen for $\hat{\mathrm{A}}$.
In the continuum SymEFT we use dimensional regularisation combined with the $\MSbar$ renormalisation scheme~\cite{tHooft:1972tcz,tHooft:1973mfk,Bardeen:1978yd}, here indicated by the subscripts used on the renormalised operators.
\Eq{eq:fpiExample} assumes that the renormalisation scheme chosen on the lattice does not introduce additional lattice artifacts.
In general this is not the case and the renormalisation condition must be taken into account as a source of lattice artifacts as well.
Including these effects does in principle not pose a problem and only doubles the non-trivial terms in \eq{eq:fpiExample}.
The generalisation to multiple and different local fields is straight forward.

The contribution from the effective action will inevitably introduce contact terms with the local fields.
Renormalisation of those contact terms can be absorbed into the renormalisation of the higher-dimensional local fields $\Jop^{(d)}$ in the continuum theory~\cite{Luscher:1996sc}.
Notice that this will impact the matching coefficients $d_i^J$ already at tree-level as we will see in section~\ref{sec:contactTermRenorm}.
Let us stress again, that we have explicitly excluded the case of having contact terms in the lattice theory, e.g., when the local fields are integrated over some region in space-time that overlaps with another local field.
Whether and how SymEFT can then still be applied after the contact interactions have been properly renormalised in the lattice theory, is an issue of its own and would go beyond the scope of this paper.
For ideas on how to treat some of those cases see~\cite{Ce:2021xgd,Sommer:2022wac}.

%% file: minimalBasis.tex
\section{Minimal operator bases to mass-dimension~5}\label{sec:minBasis}
We consider the local gauge-invariant fields of the form
\begin{equation}
\Jop^{kl}(x)=[\bar{q}_k \Gamma q_l](x),
\end{equation}
where $\bar{q}_k$ and $q_l$ are (possibly differently) flavoured quarks, and Euclidean $\Gamma\in\{1,\allowbreak\gamma_5,\allowbreak\gamma_\mu,\allowbreak\gamma_\mu\gamma_5,\allowbreak i\sigma_{\mu\nu}\}$ corresponding to the fermion bilinears $J\in \{\mathrm{S}, \mathrm{P}, \mathrm{V}, \mathrm{A}, \mathrm{T}\}$, namely the scalar~($\mathrm{S}$), pseudo-scalar~($\mathrm{P}$), vector~($\mathrm{V}$), axial-vector~($\mathrm{A}$) and tensor~($\mathrm{T}$) respectively.
Here $\sigma_{\mu\nu}=i[\gamma_\mu,\gamma_\nu]/2$.
As was the case for the lattice artifacts originating from the lattice action, the operators describing the lattice artifacts from local fields are severely constrained by the transformation properties of the local field on the lattice except for those transformations already broken by the lattice action.
We will assume here the use of either Wilson or GW quarks in the sea and valence sector, but leave the freedom of having different discretisations in both sectors.
This enforces graded $\mathrm{SU}(\Nf)_\mathrm{V}\times\mathrm{SU}(\Nb|\Nb)_\mathrm{V}$ flavour symmetry on the massless mixed action \cite{Bar:2003mh,Husung:2022kvi}, which in turn limits the operator bases of both the action and bilinear.
Here $\Nb$ is the number of (quenched) valence quarks.
For bilinears this choice affects primarily which massive operators are allowed to occur at $\ord(a)$.
In case the discretisation agrees in both sectors, the flavour symmetries in the massless limit become more stringent as $\mathrm{SU}(\Nf+\Nb|\Nb)_\mathrm{V}$.

Further limiting ourselves to lattice artifacts of at most $\ord(a^2)$, parity and time reflections reduce the set of operator candidates to a few fermion bilinears and purely gluonic operators of at most mass-dimension~5.
Furthermore, the transformation behaviour under charge conjugation combined with the graded $\mathrm{SU}(\Nb|\Nb)_\mathrm{V}$ flavour symmetry in the massless limit requires
\begin{equation}
\Jop^{kl}\stackrel{\C}{\longrightarrow} \eta_\C \Jop^{lk},
\end{equation}
where $\eta_\C=\pm 1$ depending on the Dirac matrix $\Gamma$.
This constraint requires covariant derivatives to act on both flavours equally up to a relative sign depending on $\eta_\C$.
The required transformation properties under parity, time reversal and charge conjugation are listed in table~\ref{tab:transfProperties}.

\begin{table}\setlength{\tabcolsep}{4pt}
\caption{Transformation properties of the local fields $\Jop^{kl}(x)$.
We include both parity and time reversal $\Jop^{kl}\stackrel{\P,\T}{\longrightarrow}\eta_{\P,\T}\Jop^{kl}$ as well as charge conjugation $\Jop^{kl}\stackrel{\C}{\longrightarrow} \eta_\C \Jop^{lk}$ for the scalar~(S), pseudo-scalar~(P), vector~($\mathrm{V}_\mu$), axial-vector~($\mathrm{A}_\mu$) and tensor~($\mathrm{T}_{\mu\nu}$).
To help elucidate why only specific operators are allowed up to mass-dimension~5, we include the derivative and the (dual) field-strength tensor.}\label{tab:transfProperties}\centering
\begin{tabular}{c|ccccc|ccc}
          &  S  &  P  & $\mathrm{V}_\mu$         & $\mathrm{A}_\mu$         & $\mathrm{T}_{\mu\nu}$                & $\partial_\mu$           &  $F_{\mu\nu}$                         & $\tilde{F}_{\mu\nu}$\\[3pt]\hline\hline
$\eta_\C$ & $+$ & $+$ & $-$                      & $+$                      & $-$                                  & $+$                      & $-$                                  & $-$ \\
$\eta_\P$ & $+$ & $-$ & $(-1)^{\delta_{\mu0}-1}$ & $(-1)^{\delta_{\mu0}}$   & $(-1)^{\delta_{\mu0}+\delta_{\nu0}}$ & $(-1)^{\delta_{\mu0}-1}$ & $(-1)^{\delta_{\mu0}+\delta_{\nu0}}$ & $(-1)^{\delta_{\mu0}+\delta_{\nu0}-1}$ \\
$\eta_\T$ & $+$ & $-$ & $(-1)^{\delta_{\mu0}}$   & $(-1)^{\delta_{\mu0}-1}$ & $(-1)^{\delta_{\mu0}+\delta_{\nu0}}$ & $(-1)^{\delta_{\mu0}}$   & $(-1)^{\delta_{\mu0}+\delta_{\nu0}}$ & $(-1)^{\delta_{\mu0}+\delta_{\nu0}-1}$
\end{tabular}
\end{table}
Finally, we take into account what happens under a global phase transformation of a single flavour
\begin{equation}
\bar{q}_k\rightarrow e^{-i\varphi}\bar{q}_k,\quad q_k\rightarrow e^{i\varphi}q_k,\quad \varphi=\text{const.},\quad \text{$k$ fixed},
\end{equation}
which leads to
\begin{equation}
J^{kk}(x)\rightarrow J^{kk}(x),\quad \Jop^{kl}(x)\rightarrow e^{-i\varphi}\Jop^{kl}(x),\quad \Jop^{lk}(x)\rightarrow e^{i\varphi}\Jop^{lk}(x),\quad l\neq k.
\end{equation}
Unless the lattice action introduces flavour-changing interactions\footnote{Neglecting this case explicitly excludes staggered quarks~\cite{Kogut:1974ag} from our analysis because of the presence of flavour-changing interactions in this formulation of lattice QCD.}we conclude that fermion bilinears mix only with quark anti-quark pairs of identical flavours, and purely-gluonic operators are relevant only for bilinears with trivial flavour quantum numbers, i.e., $\Jop^{kl}|_{l=k}$.

Combining these constraints allows us to list the minimal on-shell operator basis to mass-dimension~5 for the scalar (excluding the case with trivial flavour quantum numbers),
\begingroup\allowdisplaybreaks
\begin{subequations}\label{eq:scalarBasis}
\begin{align}
\S^{(1)}_1&=\frac{m_{k+l}}{2}\mathrm{S}^{k\neq l},&
\S^{(1)}_2&=\tr(M)\mathrm{S}^{k\neq l},\\
\S^{(2)}_1&=\frac{i}{4}\bar{q}_k\sigma_{\kappa\lambda}F_{\kappa\lambda}q_l,&
\S^{(2)}_2&=\partial^2\mathrm{S}^{k\neq l},\nonumber\\*
\S^{(2)}_3&=m_{k-l}^2\mathrm{S}^{k\neq l},&
\S^{(2)}_{4}&=\frac{m_{k+l}^2}{4}\mathrm{S}^{k\neq l},\nonumber\\*
\S^{(2)}_{4+j}&=\tr(M)\S^{(1)}_j,&
\S^{(2)}_{7}&=\tr(M^2)\mathrm{S}^{k\neq l},
\end{align}
\end{subequations}
where we introduced the sloppy shorthand $m_{k\pm l}=m_k\pm m_l$ with masses $m_k$ and $m_l$ of the corresponding quark flavours.
Analogously, we find for the pseudo-scalar
\begin{subequations}\label{eq:pseudoscalarBasis}
\begin{align}
\PS^{(1)}_1&=\frac{\delta_{kl}}{\bare{g}^2}\tr(F_{\rho\lambda}\tilde{F}_{\rho\lambda}),&
\PS^{(1)}_2&=\frac{m_{k+l}}{2}\mathrm{P}^{kl},\nonumber\\*
\PS^{(1)}_3&=\tr(M)\mathrm{P}^{kl},\\
\PS^{(2)}_1&=\frac{i}{4}\bar{q}_k\sigma_{\rho\lambda}\tilde{F}_{\rho\lambda}q_l,&
\PS^{(2)}_2&=\partial^2\mathrm{P}^{kl},\nonumber\\*
\PS^{(2)}_3&=m_{k-l}^2\mathrm{P}^{kl},&
\PS^{(2)}_{3+(j<3)}&=\frac{m_{k+l}}{2}\PS^{(1)}_j,&\nonumber\\*
\PS^{(2)}_{5+j}&=\tr(M)\PS^{(1)}_j,&
\PS^{(2)}_{9}&=\tr(M^2)\mathrm{P}^{kl},
\end{align}
\end{subequations}
with the dual field-strength tensor $\tilde{F}_{\mu\nu}=-\varepsilon_{\mu\nu\rho\sigma}F_{\rho\sigma}/2$
and for the vector
\begin{subequations}\label{eq:vectorBasis}
\begin{align}
\V^{(1)}_1&=\partial_\nu \mathrm{T}_{\nu\mu}^{kl},&
\V^{(1)}_2&=\frac{m_{k+l}}{2}\mathrm{V}_\mu^{kl},\nonumber\\*
\V^{(1)}_3&=\tr(M)\mathrm{V}_\mu^{kl},\\
\V^{(2)}_1&=\delta_{\mu\rho\lambda}\bar{q}_k\gamma_\rho\cevvec{D^{\mathrlap{\smash{2}}}_{\lambda}}q_l,&
\V^{(2)}_2&=i\bar{q}_k \gamma_{\rho}\gamma_5\tilde{F}_{\rho\mu}q_l,\nonumber\\*
\V^{(2)}_3&=\delta_{\mu\rho\lambda}\partial_\rho^2\mathrm{V}_\lambda^{kl},&
\V^{(2)}_4&=\partial^2\mathrm{V}_\mu^{kl},\nonumber\\*
\V^{(2)}_5&=m_{k-l}^2\mathrm{V}_\mu^{kl},&
\V^{(2)}_6&=m_{k-l}\partial_\mu\mathrm{S}^{k\neq l},&\nonumber\\*
\V^{(2)}_{6+(j<3)}&=\frac{m_{k+l}}{2}\V^{(1)}_j,&
\V^{(2)}_{8+j}&=\tr(M)\V^{(1)}_j,&\nonumber\\*
\V^{(2)}_{12}&=\tr(M^2)\mathrm{V}_\mu^{kl},
\end{align}
\end{subequations}
where we introduced the shorthands $\cevvec{D^{\mathrlap{\smash{2}}}_{\lambda}}=\cev{D}{}^2_\lambda+D^2_\lambda$ and $\delta_{\mu_1\ldots\mu_n}$ being the generalisation of the Kronecker delta to $n$ indices.
For covariant derivatives acting to the right the arrow has always been omitted.
Similarly we find for the axial-vector
\begin{subequations}\label{eq:axialvectorBasis}
\begin{align}
\A^{(1)}_1&=\partial_\mu\mathrm{P}^{kl},&
\A^{(1)}_2&=\frac{m_{k+l}}{2}\mathrm{A}_\mu^{kl},\nonumber\\*
\A^{(1)}_3&=\tr(M)\mathrm{A}_\mu^{kl},\\
\A^{(2)}_1&=\delta_{\mu\rho\lambda}\bar{q}_k\gamma_5\gamma_\rho\cevvec{D^{\mathrlap{\smash{2}}}_{\lambda}}q_l,&
\A^{(2)}_2&=\bar{q}_k\gamma_\rho \tilde{F}_{\mu\rho}q_l,\nonumber\\*
\A^{(2)}_3&=m_{k-l}\bar{q}_k(\cev{D}_\mu -D_\mu)\gamma_5q_l,&
\A^{(2)}_{4}&=\frac{\delta_{kl}}{\bare{g}^2}\delta_{\mu\nu\rho\sigma}\tr(D_\nu F_{\rho\lambda}\tilde{F}_{\sigma\lambda}),\nonumber\\*
\A^{(2)}_5&=\delta_{\mu\rho\lambda}\partial_\rho^2\mathrm{A}_\lambda^{kl},&
\A^{(2)}_6&=\partial^2\mathrm{A}_\mu^{kl},\nonumber\\*
\A^{(2)}_7&=m_{k-l}^2\mathrm{A}_\mu^{kl},&
\A^{(2)}_{8}&=\partial_\mu\PS^{(1)}_1,\nonumber\\*
\A^{(2)}_{8+(j<3)}&=\frac{m_{k+l}}{2}\A^{(1)}_j,&
\A^{(2)}_{10+j}&=\tr(M)\A^{(1)}_j,\nonumber\\*
\A^{(2)}_{14}&=\tr(M^2)\mathrm{A}_\mu^{kl},
\end{align}
\end{subequations}
and for the tensor
\begin{subequations}\label{eq:tensorBasis}
\begin{align}
\TT^{(1)}_1&=\partial_\mu\mathrm{V}_\nu^{kl}-\partial_\nu\mathrm{V}_\mu^{kl},&
\TT^{(1)}_2&=\frac{m_{k+l}}{2}\mathrm{T}_{\mu\nu}^{kl},&\nonumber\\
\TT^{(1)}_3&=\tr(M)\mathrm{T}_{\mu\nu}^{kl},&\\
\TT^{(2)}_1&=i(\delta_{\mu\kappa\rho}\delta_{\nu\lambda}+\delta_{\mu\kappa}\delta_{\nu\lambda\rho})\bar{q}_k\sigma_{\kappa\lambda}\cevvec{D^{\mathrlap{\smash{2}}}_{\rho}}q_l,&
\TT^{(2)}_2&=\bar{q}_kF_{\mu\nu}q_l,\nonumber\\*
\TT^{(2)}_3&=\bar{q}_k\gamma_5\tilde{F}_{\mu\nu}q_l,&
\TT^{(2)}_4&=(\delta_{\mu\rho\kappa}\delta_{\nu\lambda}+\delta_{\mu\kappa}\delta_{\nu\rho\lambda})\partial_\rho^2\mathrm{T}_{\kappa\lambda}^{kl},\nonumber\\*
\TT^{(2)}_5&=\partial^2\mathrm{T}_{\mu\nu}^{kl},&
\TT^{(2)}_6&=\partial_\mu\partial_\rho\mathrm{T}_{\rho\nu}^{kl}-\partial_\nu\partial_\rho\mathrm{T}_{\rho\mu}^{kl},\nonumber\\*
\TT^{(2)}_7&=m_{k-l}^2\mathrm{T}_{\mu\nu}^{kl},&
\TT^{(2)}_{8}&=m_{k-l}\varepsilon_{\mu\nu\lambda\rho}\partial_\rho\mathrm{A}^{kl}_\lambda,\nonumber\\*
\TT^{(2)}_{8+(j<3)}&=\frac{m_{k+l}}{2}\TT^{(1)}_j,&
\TT^{(2)}_{10+j}&=\tr(M)\TT^{(1)}_j,\nonumber\\*
\TT^{(2)}_{14}&=\tr(M^2)\mathrm{T}_{\mu\nu}^{kl}.
\end{align}
\end{subequations}
\endgroup
Since we will be working with an \emph{off-shell} renormalisation strategy we also need to keep track of a minimal basis of \eom{} operators for each of the local fields.
Those bases are listed in the appendix in \eqs{eq:Jeom}.
In the presence of contact terms with other local fields those operators would become relevant to leading order in the lattice spacing and could no longer be ignored.

The minimal on-shell bases at mass-dimension~4 are found to be the same as the ones listed in~\cite{Luscher:1996sc,Bhattacharya:2003nd,Bhattacharya:2005rb}.
The flavour-singlet quark-bilinear operators, that are needed to renormalise the purely-gluonic flavour-singlet operators, are in general omitted.
Those are implicitly included by forming the correct linear combinations of the bases given in \eqs{eq:scalarBasis}--\eqref{eq:tensorBasis}.
Notice that there are both, a singlet for the sea sector and a quenched singlet for the valence sector.

Although we find a large number of operators for each fermion bilinear, there are not too many genuinely new operators in the sense that all massive operators and total derivative operators are in principle known.
They renormalise just like their lower-dimensional counterparts up to a multiplicative renormalisation factor for the quark masses.
We thus find only a very limited number of entirely new operators up to mass-dimension~5 relevant for each of the local fields.

\paragraph{Remarks on the derivation of the minimal basis:}
\begin{itemize}
\item All fermion bilinears that are not the original local fields themselves dressed with powers of quark masses or total derivatives thereof occur first at $\ord(a^2)$.
The only ``new'' fields at $\ord(a)$ are purely gluonic and contribute only for the case of trivial flavour quantum numbers.
\item For the case of non-trivial flavour quantum numbers massive operators must be taken into account, that have the wrong transformation behaviour under charge conjugation when stripped off their mass-difference prefactor, e.g., for the vector
\begin{equation}
\V^{(2)}_6=m_{k-l}\partial_\mu S^{k\neq l}.
\end{equation}
Only for the axial-vector an entirely new operator $\A_3^{(2)}$ becomes relevant at $\ord(a^2)$ which has peculiar quantum numbers when stripped off its mass-difference prefactor and is thus not covered in \tab{tab:transfProperties} prior to $\ord(a^2)$.
\item Since the vector and tensor are odd under charge conjugation, there are no new purely gluonic operators in contrast to, e.g., the axial-vector.
\item The reduction of the operator basis was performed with the following hierarchy in mind
\begin{equation*}
\text{EOM-vanishing}>\text{massive}>\text{total divergences}>\text{others},
\end{equation*}
where $>$ indicates which operators to keep during the reduction of the minimal basis.
Apart from the EOM-vanishing operators, which must be prioritised to work out the minimal on-shell basis, the hierarchy is totally arbitrary.
This particular choice has its merits, when going to the massless limit or the mass-degenerate case.
It may also be beneficial for massive renormalisation schemes such as, e.g.,~\cite{Boyle:2016wis,Fritzsch:2018kjg}.
\item The presence of the Levi-Civita tensor or $\gamma_5$ in general leads to complications in dimensional regularisation.
Since we aim only at the 1-loop UV-divergent part, we can ignore such subtleties.
Still some peculiarities are manifest, e.g., for the axial-vector basis we find the equivalence of the purely-gluonic operators
\begin{equation}
\frac{1}{\bare{g}^2}\partial_\mu\tr(F_{\kappa\lambda}\tilde{F}_{\kappa\lambda})=\frac{4}{\bare{g}^2}\partial_\kappa\tr(F_{\kappa\lambda}\tilde{F}_{\mu\lambda})=\frac{4}{\bare{g}^2}\delta_{\mu\nu\rho\sigma}\partial_\nu\tr(F_{\rho\kappa}\tilde{F}_{\sigma\kappa})\label{eq:pureGauge4Dequiv}
\end{equation}
in 4 dimensions.
In dimensional regularisation we thus expect at least two evanescent operators to be needed for this case.
For a discussion on evanescent operators see~\cite{Buras:1989xd}.
Those evanescent operators do not contribute to 1-loop divergences and we can thus choose any operator from the three variants given in \eq{eq:pureGauge4Dequiv}.
However, going beyond the 1-loop divergent part would certainly require an in-depth analysis.
For a discussion on how to treat the Levi-Civita tensor properly for the topological-charge density, the axial-vector, and the pseudo-scalar in an otherwise $\gamma_5$-free theory see, e.g., \cite{Luscher:2021bog} and references therein.
\item The axial-vector is the only bilinear that incorporates at $\ord(a^2)$ the gluonic EOM into its minimal basis of EOM-vanishing operators, see \eq{eq:axialEOM}.
As a consequence the $\ord(a^2)$ trivial-flavour on-shell operator basis may mix with the appropriate flavour singlets of the sea and valence.
\end{itemize}

%% file: opsRenorm.tex
\section{Renormalisation of the minimal operator bases to 1-loop}\label{sec:OpsRenorm}
\begin{figure}\centering
\begin{subfigure}[t]{0.24\textwidth}\centering
\includegraphics[]{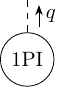}
\caption{}\label{fig:P}
\end{subfigure}
\begin{subfigure}[t]{0.24\textwidth}\centering
\includegraphics[]{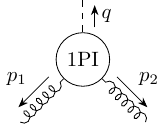}
\caption{}\label{fig:B2P}
\end{subfigure}
\begin{subfigure}[t]{0.24\textwidth}\centering
\includegraphics[]{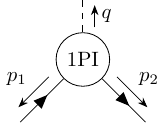}
\caption{}\label{fig:F2P}
\end{subfigure}
\begin{subfigure}[t]{0.24\textwidth}\centering
\includegraphics[]{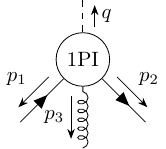}
\caption{}\label{fig:F2BP}
\end{subfigure}
\caption{1PI Feynman graphs computed to determine the renormalisation of our local fields inserted at momentum $q$ (dashed line).
The incoming and outgoing fermion lines can be of different quark flavour to allow for non-trivial flavour quantum numbers.}\label{fig:renormalisationLocalFields}
\end{figure}

The overall renormalisation strategy used here is very similar to the one discussed in~\cite{Husung:2022kvi}.
In addition to having multiple operator bases for the local fields and the SymEFT action, two key differences should be noted:
\begin{enumerate}
\item The operator insertions for the local fields are now at non-zero momentum $q$ as depicted in \fig{fig:renormalisationLocalFields}.
This figure represents all graphs necessary to perform the 1-loop renormalisation of the minimal bases given in \eqs{eq:scalarBasis}--\eqref{eq:tensorBasis}.
The graph in \fig{fig:P} would only be relevant for the scalar with trivial flavour quantum numbers.
\item Contact terms arise when the spacetime arguments of the local field and of an operator insertion from the expanded SymEFT action coincide.
These must be renormalised on top of the renormalisation of the individual composite fields.
The treatment of such contact terms holds some peculiarities and is covered in its own dedicated subsection~\ref{sec:contactTermRenorm}.
\end{enumerate}
For compactness, we discard any explicit flavour indices as well as superscripts ${}^{(d)}$ and instead discuss the renormalisation for generic local fields characterised by their quantum numbers.
Working in background-field gauge~\cite{tHooft:1975uxh,Abbott:1980hw,Abbott:1981ke,Luscher:1995vs}, the overall mixing matrix involves only (background-)gauge-invariant local fields of the same mass-dimension and takes the block form
\begin{equation}
\begin{pmatrix}
\Jop \\[6pt]
\JopE
\end{pmatrix}_{\MSbar}=\begin{pmatrix}
Z^{\Jop} & Z^{\Jop\JopE} \\[6pt]
0        & Z^{\JopE}
\end{pmatrix}
\begin{pmatrix}
\Jop\\[6pt]
\JopE
\end{pmatrix},\label{eq:triangMixing}
\end{equation}
where $Z^{\Jop}$ is the desired on-shell mixing matrix and $Z^{\Jop\JopE}$ are the mixing contributions from operators $\JopE$ vanishing by the \emph{classical} EOMs~\cite{Kluberg-Stern:1975ebk}.
They are therefore irrelevant for physical on-shell observables \emph{in the absence of contact terms}.
This vanishing by EOMs is also the reason for the triangular mixing structure.

In the following, we will give only the 1-loop anomalous-dimension matrices, i.e., $\gamma_0^\Jop$ as defined through the Renormalisation Group Equation (RGE)
\begin{equation}
\mu\frac{\rmd}{\rmd \mu}\begin{pmatrix}
\Jop \\[6pt]
\JopE
\end{pmatrix}_{\MSbar}=-\gbar^2(\mu)
\begin{pmatrix}
\gamma_0^\Jop & \gamma_0^{\Jop\JopE} \\[6pt]
0 & \gamma_0^{\JopE}
\end{pmatrix}
\begin{pmatrix}
\Jop \\[6pt]
\JopE
\end{pmatrix}_{\MSbar}
+\ord(\gbar^4(\mu))\,.\label{eq:RGE_J}
\end{equation}
Once the 1-loop anomalous-dimension matrices for the operator basis are known, we can make the usual change of basis
\begin{equation}
\begin{pmatrix}
\Jbase \\[6pt]
\JbaseE
\end{pmatrix}_{\MSbar}=
\begin{pmatrix}
T^J & T^{\Jop\JopE} \\[6pt]
0   & 1
\end{pmatrix}
\begin{pmatrix}
\Jop \\[6pt]
\JopE
\end{pmatrix}_{\MSbar}\label{eq:base2Jordan}
\end{equation}
bringing the on-shell part of the 1-loop anomalous-dimension matrix into Jordan normal form, see also the discussion in \cite{Husung:2022kvi}.
Notice, that the blocks $T^{\Jop\JopE}$ and $T^{\JopE}=1$ have only been added for consistency, as they do not affect the 1-loop anomalous dimensions of the on-shell basis.
In the absence of contact terms, we can restrict ourselves to the on-shell basis, i.e., the block matrix $T^\Jop$.

We now come back to our various bases of local fields $\Jop^{(d)}$ for $J\in\{\mathrm{S},\mathrm{P},\mathrm{V},\mathrm{A},\mathrm{T}\}$.
As a check, we first computed the 1-loop anomalous dimensions for the continuum fields
\begin{align}
(4\pi)^2\gamma_0^{\mathrm{S}^{k\neq l}}=(4\pi)^2\gamma_0^{\mathrm{P}^{kl}}=3\frac{1-\Nc^2}{\Nc},\quad \gamma_0^{\mathrm{V}^{kl}}=\gamma_0^{\mathrm{A}^{kl}}=0,\quad (4\pi)^2\gamma_0^{\mathrm{T}^{kl}}=\frac{\Nc^2-1}{\Nc}.\label{eq:dim3Mixing}
\end{align}
They agree with the values found in the literature~\cite{Larin:1993vu,Broadhurst:1994se}.

Next we give the 1-loop anomalous dimensions found, again discarding operators carrying traces of the quark-mass matrix for compactness.
For the on-shell basis at mass-dimension~4 we find genuine new operators only for the pseudo-scalar having trivial flavour quantum numbers
\begin{equation}
(4\pi)^2\AD[\mathrm{P}^{kl}]{1}=\left(\begin{array}{c;{1pt/1pt}c}
-2b_0\delta_{kl} & 6\frac{1-\Nc^2}{\Nc}\Sigma
\end{array}\right).
\end{equation}
Here and in the following, the dotted vertical line splits the mixing into contributions from genuinely new operators at the current mass-dimension to the left and those operators whose mixing can be inferred from lower-dimensional versions here being omitted from the full mixing-matrix for compactness.
To distinguish mixing contributions that are only present for trivial flavour quantum numbers we further introduce a Kronecker $\delta_{kl}$ for contributions from purely gluonic operators and $\Sigma$ to indicate the summation over all flavours from sea and valence.
The overall ordering of the rows and columns is always according to \eqs{eq:scalarBasis}--\eqref{eq:tensorBasis}.

Operators whose mixing can be inferred from their lower-dimensional versions are total divergences of the local field itself or carry overall powers of quark masses.
For the latter case with arbitrary integer powers $n$ of quark masses, we can always use
\begin{equation}
\gamma_0^{m^nJ}=n \gamma_0^m+\gamma_0^J\,,\quad (4\pi)^2\gamma_0^m=3 \frac{\Nc^2-1}{\Nc}\,,
\end{equation}
where $\gamma_0^m$ denotes the 1-loop anomalous dimension of the quark mass.
The same holds for total divergences, but there are some peculiarities where, e.g., initially axial-like operators mix into tensor-like operators with a proper contraction of Lorentz indices.
For compactness we discard all operators carrying some trace of the sea-quark mass-matrix.

At mass-dimension~5 we then find new operators with 1-loop anomalous dimensions
\begingroup\allowdisplaybreaks
\begin{align}
(4\pi)^2\AD[\mathrm{S}^{k\neq l}]{2}&=\left(\arraycolsep=2pt\begin{array}{c;{1pt/1pt}ccc}
\frac{\Nc^2-5}{\Nc} & 0 & \frac{3-3\Nc^2}{4\Nc} & \frac{3-3\Nc^2}{\Nc}
\end{array}\right),\nonumber\\[4pt]
(4\pi)^2\AD[\mathrm{P}^{k l}]{2}&=\left(\arraycolsep=2pt\begin{array}{c;{1pt/1pt}cccc}
\frac{\Nc^2-5}{\Nc} & 0 & \frac{3-3\Nc^2}{4\Nc} & 2\delta_{kl} & \frac{3-3\Nc^2}{\Nc}
\end{array}\right),\nonumber\\[4pt]
\frac{(4\pi)^2\Nc}{\Nc^2-1}\AD[\mathrm{V}^{k l}]{2}&=\left(\arraycolsep=2pt\begin{array}{cc;{1pt/1pt}cccccc}
\frac{25}{6} & \frac{1}{2} & -\frac{5}{2} & \frac{5}{12} & -\frac{1}{8} & \frac{1}{4} & \frac{1}{6} & \frac{7}{6} \\[6pt]
0 & \frac{8}{3} & 0 & 0 & \frac{5}{6} & -\frac{1}{3} & -\frac{2}{3} & 2
\end{array}\right),\nonumber\\[4pt]
\frac{(4\pi)^2\Nc}{\Nc^2-1}\AD[\mathrm{A}^{k l}]{2}&=\left(\arraycolsep=2pt\begin{array}{cccc;{1pt/1pt}cccccc}
 \frac{25}{6} & \frac{3-3\Nc^2+4 \Nc \Sigma}{6-6 \Nc^2} & \frac{1}{12} & \frac{8 \Nc\delta_{kl}}{3-3 \Nc^2} & -\frac{5}{2} & \frac{5}{12} & \frac{5}{24} & \frac{\Nc\delta_{kl}}{3 \left(\Nc^2-1\right)} & \frac{1}{2} & -\frac{1}{2} \\[6pt]
 0 & \frac{8-8 \Nc^2-4 \Nc \Sigma}{3-3 \Nc^2} & -\frac{1}{3} & 0 & 0 & 0 & \frac{5}{6} & 0 & -\frac{2}{3} & \frac{10}{3} \\[6pt]
 0 & 0 & 4 & 0 & 0 & 0 & 2 & 0 & 0 & 0 \\[6pt]
 -\frac{5\Sigma}{24} & \frac{\left(5 \Nc^2+7\right) \Sigma}{24 \left(\Nc^2-1\right)} & 0 & \frac{\Nc \left(18 \Nc+4 \Nf\right)}{3\Nc^2-3}\delta_{kl} & \frac{\Sigma}{8} & -\frac{\Sigma}{48} & 0 & \frac{5 \Nc^2\delta_{kl}}{3-3 \Nc^2} & -\frac{17\Sigma}{24} & -\frac{5\Sigma}{24}
\end{array}
\right),\nonumber\\[4pt]
\frac{(4\pi)^2\Nc}{\Nc^2-1}\AD[\mathrm{T}^{kl}]{2}&=\left(\arraycolsep=2pt\begin{array}{ccc;{1pt/1pt}ccccccc}
 \frac{13}{3} & \frac{2 \Nc^2+28}{3-3 \Nc^2} & -\frac{2 \left(\Nc^2+8\right)}{3 \left(\Nc^2-1\right)} & -2 & \frac{2}{3} & \frac{2}{3} & \frac{4}{3} & \frac{2}{3} & \frac{2}{3} & 0 \\[6pt]
 0 & \frac{7 \left(2 \Nc^2+1\right)}{3 \left(\Nc^2-1\right)} & \frac{\Nc^2+2}{3 \left(\Nc^2-1\right)} & 0 & 0 & 0 & \frac{1}{3} & \frac{1}{6} & -\frac{1}{3} & \frac{2}{3} \\[6pt]
 0 & \frac{\Nc^2+2}{3 \left(\Nc^2-1\right)} & \frac{7 \left(2 \Nc^2+1\right)}{3 \left(\Nc^2-1\right)} & 0 & 0 & 0 & \frac{1}{6} & -\frac{1}{6} & \frac{1}{3} & \frac{4}{3}
\end{array}\right).
\end{align}
\endgroup\allowdisplaybreaks
No explicit matrix is given for the singlets for the sea or valence respectively as their form can be inferred from the details given here with some effort or simply obtained from the supplemental material.
Keep in mind that the valence-singlets are quenched.

After making the change of basis as indicated in \eq{eq:base2Jordan}, we can finally determine the leading powers in $\gbar^2(1/a)$ introduced by the leading lattice artifacts $\Jbase^{(d)}$ of the local field~$\Jop$ via renormalisation group running, see \cite{Papinutto:2016xpq} as well as \cite{Husung:2022kvi},
\begin{equation}
\hat{\gamma}^\Jop\define\frac{\diag\left(\AD[\Jbase]{d}\right)-\gamma_0^\Jop}{2b_0}\label{eq:gammaJ}
\end{equation}
up to positive integer-shifts due to vanishing tree-level matching coefficients.
The subtraction of $\gamma_0^\Jop$ occurs due to the normalisation with the continuum counterpart as can e.g.~be seen in \eq{eq:fpiExample}.
At this point it should be noted, that $\hat{\gamma}^\Jop$ itself does not depend on any contact-term renormalisation.
If we were only interested in those powers, we could stop here.
As we will see in \sect{sec:contactTermRenorm}, contact terms with the SymEFT action may further modify the leading asymptotic lattice spacing dependence by factoring in an additional $\log(2b_0\gbar^2(1/a))$, i.e., $[2b_0\gbar^2(1/a)]^{\hat{\Gamma}_i}\rightarrow [2b_0\gbar^2(1/a)]^{\hat{\Gamma}_i}\log(2b_0\gbar^2(1/a))$.

As an intermediate result we list the $\hat{\gamma}^\Jop$ found in table~\ref{tab:1loopAD} for some choices of $\Nf$.
Only the flavour-neutral axial-vector converges (significantly) slower than classical $a^2$ for the fully $\ord(a)$-improved case while otherwise only the pseudo-scalar has a slightly negative powers, again for the flavour-neutral case.
Without explicit $\ord(a)$ improvement, the situation is more complicated because both axial-vector and tensor have negative powers for arbitrary flavours, while the pseudo-scalar has negative powers only for the flavour-neutral case.
In finite volume with non-trivial flavour quantum numbers the $\ord(a)$ terms are suppressed by one power in the quark mass.
However, taking the axial-axial 2-point function at $\Nf=3$ as an example, both local fields will give rise to an insertion of $\ord(a)$-terms which then lead to significantly enlarged $\ord(a^2)$ effects of the asymptotically leading form $a^2[2b_0\gbar^2(1/a)]^{-0.828}$.
Fortunately, the commonly used strictly-local discretisation of the axial-vector has TL-suppressed matching coefficients for this particular contribution.


\begin{table}
\caption{Non-exhaustive examples of (distinct) 1-loop anomalous dimensions found for the local fields at $\Nf=2,3$ flavours in 3-colour lattice QCD rounded to the third decimal.
Keep in mind that $\hat{\gamma}^\Jop+1$ have not been added but will arise from loop-suppressed contributions.
\uwave{Underwiggled} numbers occur only for local fields with trivial flavour quantum numbers, i.e., $k=l$.
\underline{Underlined} numbers belong to massive contributions.
\dotuline{Underdotted} numbers correspond to massive contributions, that vanish in the mass-degenerate case and for $k=l$.
}\label{tab:1loopAD}\centering
\begin{tabular}{l|rl}
$(\Jbase^{kl})^{(1)}$   & $\Nf$ & $\hat{\gamma}^\Jop$ \\\hline
scalar ($k\neq l$ only) & 2     &  \underline{$0.414$} \\
                        & 3     &  \underline{$0.444$} \\
pseudo-scalar           & 2     &  \uwave{$-0.586$}, \underline{$0.414$} \\
                        & 3     &  \uwave{$-0.556$}, \underline{$0.444$} \\
vector                  & 2     &  $0.138$, \underline{$0.414$} \\
                        & 3     &  $0.148$, \underline{$0.444$} \\
axial-vector            & 2     &  $-0.414$, \underline{$0.414$} \\
                        & 3     &  $-0.444$, \underline{$0.444$}\\
tensor                  & 2     &  $-0.138$, \underline{$0.414$} \\
                        & 3     &  $-0.148$, \underline{$0.444$} \\\hline\hline\\[-12pt]
$(\Jbase^{kl})^{(2)}$   & $\Nf$ & $\hat{\gamma}^\Jop$ \\\hline
scalar ($k\neq l$ only) & 2     & $0$, $0.483$, \underline{$0.828$} \\
                        & 3     & $0$, $0.519$, \underline{$0.889$} \\
pseudo-scalar           & 2     & \underline{\uwave{$-0.172$}}, $0$, $0.483$, \underline{$0.828$} \\
                        & 3     & \underline{\uwave{$-0.111$}}, $0$, $0.519$, \underline{$0.889$} \\
vector                  & 2     & $0$, $0.368$, \underline{$0.552$}, $0.575$, \underline{$0.828$} \\
                        & 3     & $0$, $0.395$, \underline{$0.593$}, $0.617$, \underline{$0.889$} \\
axial-vector            & 2     & \uwave{$-1$}, $0$, $0.368$, \uwave{$0.506$}, \dotuline{$0.552$}, \uwave{$0.559$}, $0.575$, \underline{$0.828$},  \uwave{$1.085$}  \\
                        & 3     & \uwave{$-1$}, $0$, $0.395$, \dotuline{$0.593$}, \uwave{$0.595$}, $0.617$, \underline{$0.889$},  \uwave{$1.244$} \\
tensor                  & 2     & $0$, \underline{$0.276$}, $0.46$, $0.563$, $0.69$, \underline{$0.828$}\\
                        & 3     & $0$, \underline{$0.296$}, $0.494$, $0.605$, $0.741$, \underline{$0.889$}
\end{tabular}
\end{table}

\subsection{Contact terms of local fields with operators from the effective action}\label{sec:contactTermRenorm}
Finally, to obtain the tree-level matching coefficients as well, we will have to deal with the contact terms.
In perturbation theory, contact-divergences arise by construction as loop corrections, but bringing the 1-loop mixing matrix to Jordan normal form will inevitably impact the tree-level matching coefficients of some operators of the final basis $\Jbase_i^{(d)}$.
We therefore need to know the full mixing at 1-loop order to have all information to determine the tree-level matching coefficients for the final basis.

\begin{figure}\centering
\begin{subfigure}[t]{0.24\textwidth}\centering
\includegraphics[]{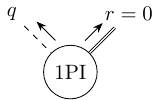}
\caption{}\label{fig:OP}
\end{subfigure}
\begin{subfigure}[t]{0.24\textwidth}\centering
\includegraphics[]{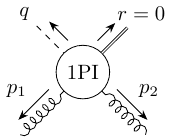}
\caption{}\label{fig:B2OP}
\end{subfigure}
\begin{subfigure}[t]{0.24\textwidth}\centering
\includegraphics[]{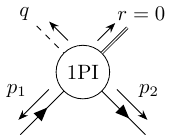}
\caption{}\label{fig:F2OP}
\end{subfigure}
\begin{subfigure}[t]{0.24\textwidth}\centering
\includegraphics[]{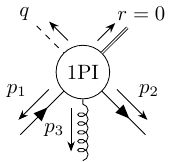}
\caption{}\label{fig:F2BOP}
\end{subfigure}
\caption{1PI Feynman graphs computed to determine the renormalisation of contact-divergences arising from the contact interaction of our local fields inserted at momentum $q$ (dashed line) with operators of the minimal basis describing lattice artifacts of the lattice action inserted at momentum $r=0$ (double line).
The incoming and outgoing fermion lines can be of different quark flavour to allow for non-trivial flavour quantum numbers.}\label{fig:contactTerms}
\end{figure}

Since we are interested only in the leading order lattice artifacts, we can restrict here all considerations to 1PI $n$-point functions with insertions of both a continuum fermion bilinear $\Jop\in \{\mathrm{S}, \mathrm{P}, \mathrm{V}, \mathrm{A}, \mathrm{T}\}$ at non-zero four-momentum and an operator $\Qop_i^{(d)}$ from the effective action at zero momentum.
All other possible contact divergences will be accompanied by higher powers in the lattice spacing during the expansion of the SymEFT and their impact is therefore expected to be suppressed in the asymptotic region.
As pointed out before, the presence of contact interactions means that we can no longer restrict considerations to the minimal on-shell basis $\op^{(d)}$ but need to take a minimal basis of \eom{} operators $\opE^{(d)}$ into account as well.
We thus consider the enlarged minimal operator basis $Q^{(d)}=\op^{(d)}\cup \opE^{(d)}$ for the contact-divergence renormalisation and find
\begin{align}
\left.\left\langle F(p) \tilde{\Jop}(-p)\tilde{\Qop}_i^{(d)}(0)\right\rangle\right|_\mathrm{1PI}^{\substack{\text{1-loop}\hfill\\\text{UV poles}}}&=
-\left.\left\langle F(p) \left\{Z^{\Qop\Jop}_{ij}\tilde\Jop_j^{(d)}(-p)+Z^{\Qop\JopE}_{ij}\tilde\JopE_j^{(d)}(-p)\right\}\right\rangle\right|_\mathrm{1PI}^\text{tree}\nonumber\\
&\hphantom{==}\times\left\{1+\ord(\gbar^2)\right\}.\label{eq:contactTermCounterTerm}
\end{align}
Here, $F(p)$ is any combination of fundamental fields, namely (anti-)quarks and background fields, carrying the overall momentum $p$.
$ Z_{ij}^{\Qop\Jop}$ and $ Z_{ij}^{\Qop\JopE}$ are the appropriate renormalisation factors to renormalise the contact divergences to 1-loop order.
The choices for the 1PI $n$-point functions with operator insertions are depicted in \fig{fig:contactTerms}.
Notice that \fig{fig:OP} would only be needed for the flavour-singlet scalar, which we explicitly excluded.
Fortunately, we do not need to include any fermion-4-point functions since the operators required to renormalise the contact-divergences are exactly the mass-dimension four and five operators from section~\ref{sec:minBasis}, which have both the correct quantum numbers and canonical mass-dimension.\footnote{Which canonical mass-dimension is needed depends on the mass-dimension of $\Qop^{(d)}$ or in other words, it depends on whether we consider an $\ord(a)$ improved action.
Since our operator basis for the SymEFT action consists of flavour-singlet scalars, the quantum numbers of the operators required to renormalise the contact interactions are entirely dictated by the local field $\Jop$ involved.
The situation will become more complicated for mixed actions or more exotic choices.}

The full renormalisation prescription including contact-divergence renormalisation can then be written in a sloppy way as
\begin{align}
\begin{pmatrix}
\Jbase^{(d)}(x) \\[3pt]
\tilde\op^{(d)}(0)\Jop(x) \\[3pt]
\opEtilde^{(d)}(0)\Jop(x)
\end{pmatrix}_{\MSbar}&=
\begin{pmatrix}
Z^{\Jbase} & 0 & 0 \\[3pt]
Z^{\op\Jbase} & Z^{\op}Z^J & Z^{\op\opE}Z^J \\[3pt]
Z^{\opE\Jbase} & 0 & Z^{\opE} Z^J
\end{pmatrix}
\begin{pmatrix}
\Jbase^{(d)}(x) \\[3pt]
\tilde\op^{(d)}(0)\Jop(x) \\[3pt]
\opEtilde^{(d)}(0)\Jop(x) \\[3pt]
\end{pmatrix},\label{eq:contactTermRenorm}
\end{align}
where $Z^{\op}$ is the mixing matrix renormalising the minimal on-shell basis of the SymEFT action we found before~\cite{Husung:2021mfl,Husung:2022kvi}, $Z^{\opE}$ is the analogue for the minimal basis of \eom{} operators, $Z^{\Jop}$ renormalises the continuum local field, $Z^{\Jbase}$ renormalises the diagonalised higher-dimensional basis for the local fields and the off-diagonal blocks correspond to the respective mixing.
In contrast to spectral quantities, we can no longer ignore the presence of the \eom{} operators $\opE^{(d)}$ as those operators will resurface in contact-interactions with the local fields.
Bear in mind that we immediately switched to the basis $\Jbase^{(d)}$ and dropped the superscript ${}^{(d)}$ from the block matrices.
Also, the extended basis of \eom{} local fields $\JbaseE^{(d)}$ was dropped from the mixing as it enlarges the mixing matrix further, while playing no role here for on-shell physics at leading order in the lattice spacing.
In practice, we will work in the basis $\Jop^{(d)}$ instead and of course need to keep track of $\JopE^{(d)}$ as well to work out the full off-shell mixing including contact terms from off-shell $n$-point functions with operator insertions from the SymEFT action, where $\JopE^{(d)}$ will be required as counter-terms.
Only at the very end, we will make the change of basis in \eq{eq:base2Jordan}.

The way we introduced the contact-divergence renormalisation is somewhat arbitrary, but allows us to treat both the continuum field and the operators of the SymEFT action as if there were no contact terms.
Thus, we can follow the usual strategy for multiplicatively renormalisable local fields as discussed in detail in~\cite{Husung:2022kvi}.
We begin by bringing the block matrix $Z^{\op}$ into Jordan normal form, while taking care of any mixing $Z^{\op\opE}$ via a change of basis analogously to \eq{eq:base2Jordan}
\begin{equation}
\begin{pmatrix}
\base \\
\baseE
\end{pmatrix}_{\MSbar}=
\begin{pmatrix}
T^\op & T^{\op\opE}\\
0     & 1
\end{pmatrix}
\begin{pmatrix}
\op \\
\opE
\end{pmatrix}_{\MSbar}.\label{eq:actionBase2Jordan}
\end{equation}
Most importantly, this can be done independently of any other off-diagonal block matrix, a reflection of the fact that spectral quantities remain unaffected by anything we discuss here.
Notice that while the \eom{} operators remain unaffected by this change of basis, they still get a new symbol to identify the corresponding matching coefficients unambiguously.
So far, we have not done anything that has not been done before for the case of spectral quantities.
After the change of basis we could safely ignore the remaining contact terms of \eom{} operators with the local field as those \eom{} operators will be absorbed into the matching as discussed in \sect{sec:TLmatching}.
Instead we will keep that mixing for now to make the effect easier to follow through to the final matching.
Now, only the off-diagonal block matrices relevant for the contact-divergence renormalisation remain
\begin{equation}
\begin{pmatrix}
\Jbase^{(d)}(x) \\[3pt]
\tilde\base^{(d)}(0)\Jop(x) \\[3pt]
\baseEtilde^{(d)}(0)\Jop(x)
\end{pmatrix}_{\MSbar}=
\begin{pmatrix}
Z^{\Jbase}       & 0            & 0 \\[3pt]
Z^{\base\Jbase}  & Z^{\base}Z^J & 0 \\[3pt]
Z^{\baseE\Jbase} & 0            & Z^{\baseE} Z^J 
\end{pmatrix}
\begin{pmatrix}
\Jbase^{(d)}(x) \\[3pt]
\tilde\base^{(d)}(0)\Jop(x) \\[3pt]
\baseEtilde^{(d)}(0)\Jop(x) 
\end{pmatrix}.\label{eq:usualMixing}
\end{equation}
To remove this remnant mixing, we make a last change of basis
\begin{equation}
\begin{pmatrix}
\Jbase^{(d)}(x) \\[3pt]
[\tilde{\base}^{(d)}(0)\Jop(x)]_{\ctdiv} \\[3pt]
[\baseEtilde^{(d)}(0)\Jop(x)]_{\ctdiv}
\end{pmatrix}_{\MSbar}=
\begin{pmatrix}
1                & 0 & 0 \\[3pt]
T^{\base\Jbase}  & 1 & 0 \\[3pt]
T^{\baseE\Jbase} & 0 & 1
\end{pmatrix}
\begin{pmatrix}
\Jbase^{(d)}(x) \\[3pt]
\tilde{\base}^{(d)}(0)\Jop(x) \\[3pt]
\baseEtilde^{(d)}(0)\Jop(x)
\end{pmatrix}_{\MSbar},\label{eq:ctdivSubtraction}
\end{equation}
where $[\ldots]_{\ctdiv}$ implies that contact divergences have been subtracted as indicated here.
These steps combined allow us to bring the (on-shell part of the) mixing matrix in \eq{eq:contactTermRenorm} at 1-loop into diagonal form or at least into Jordan normal form if it is non-diagonalisable.

For brevity we ignore the case of a double operator insertion of 
$\ord(a)$ operators from the SymEFT action as would be relevant, e.g., for twisted-mass QCD without clover improvement, i.e., relying on automatic $\ord(a)$ improvement at maximal chiral twist~\cite{Aoki:2006gh,Sint:2007ug}.
Obviously, this would complicate the situation even further.
Despite having now multiple contact terms to handle simultaneously the general strategy would remain the same.
Contrary, any remnant EOM-vanishing terms in the effective description of our local fields at $\ord(a)$ can be ignored until $\ord(a^3)$ if the lattice action is Symanzik $\ord(a)$ improved.
This is due to the absence of contact terms with other $\ord(a)$ terms from the SymEFT action that affect on-shell contributions.

The resulting block matrices for the full 1-loop anomalous-dimension matrix at mass-dimension~4 are
\begin{subequations}\label{eq:aContactTerms}
\begin{align}
(4\pi)^2\left[\gamma_0^{\op\,\mathrm{S}^{k\neq l}}\right]^{(1)}&=\left(
\begin{array}{c}
\frac{15}{2} \frac{\Nc^2-1}{\Nc}
\end{array}
\right),&
(4\pi)^2\left[\gamma_0^{\op\,\mathrm{P}^{kl}}\right]^{(1)}&=\left(
\begin{array}{cc}
 2\delta_{kl} & \frac{3}{2} \frac{\Nc^2-1}{\Nc}
\end{array}
\right),\nonumber\\
(4\pi)^2\left[\gamma_0^{\op\,\mathrm{V}^{kl}}\right]^{(1)}&=\left(
\begin{array}{cc}
 2\frac{\Nc^2-1}{\Nc} & \frac{3}{2}\frac{\Nc^2-1}{\Nc}
\end{array}
\right),&
(4\pi)^2\left[\gamma_0^{\op\,\mathrm{A}^{kl}}\right]^{(1)}&=\left(
\begin{array}{cc}
 0 & \frac{3}{2}\frac{\Nc^2-1}{\Nc}
\end{array}
\right),\nonumber\\
(4\pi)^2\left[\gamma_0^{\op\,\mathrm{T}^{kl}}\right]^{(1)}&=\left(
\begin{array}{cc}
 \frac{\Nc^2-1}{\Nc} & \frac{1-\Nc^2}{2\Nc}
\end{array}
\right),\\
(4\pi)^2\left[\gamma_0^{\opE\,\Jop^{kl}}\right]^{(1)}&=0\,\forall J\in\{\mathrm{S},\mathrm{P},\mathrm{V},\mathrm{A},\mathrm{T}\}\,.
\end{align}
\end{subequations}
We dropped again explicitly massive operators of the SymEFT action, whose contact terms can be inferred from their lower-dimensional counterparts without the explicit masses.
The operators $\op$ are either gluonic or involve valence quarks.
The ordering is as in \sects{sec:onshellActionBasis} and \ref{sec:EOMmixing}.
Contact terms of $\frac{1}{g_0^2}\tr(F_{\mu\nu}F_{\mu\nu})$ with the local fermion bilinear $J$ simply yield twice the anomalous dimensions from \eq{eq:dim3Mixing} as their off-diagonal entries due to our chosen normalisation.
Due to their size, the 1-loop anomalous-dimension block matrices for mass-dimension~5 can be found in appendix~\sect{sec:blockMatricesContact}.

%% file: matching.tex
\section{(Tree-level) matching}\label{sec:TLmatching}
To understand why we were carrying the \eom{} operators with us in section~\ref{sec:OpsRenorm}, it is instructive to take a look at the effect the changes of bases in \eqs{eq:base2Jordan}, \eqref{eq:actionBase2Jordan}, and \eqref{eq:ctdivSubtraction} have on the tree-level matching coefficients
\begin{equation}
\begin{pmatrix}
[d^{\Jbase}]_{\ctdiv} \\[3pt]
-c^{\base} \\[3pt]
-c^{\baseE}
\end{pmatrix}=\begin{pmatrix}
d^{J} \\[6pt]
-c^{\op} \\[6pt]
-c^{\opE}
\end{pmatrix}\begin{pmatrix}
T^J & 0     & 0 \\[3pt]
0   & T^\op & T^{\op\opE} \\[3pt]
0   & 0     & 1
\end{pmatrix}^{-1}\begin{pmatrix}
1                & 0 & 0 \\[3pt]
T^{\base\Jbase}  & 1 & 0 \\[3pt]
T^{\baseE\Jbase} & 0 & 1
\end{pmatrix}^{-1}.
\end{equation}
The relative signs in front of the matching coefficients of the action and the local field when expanding the SymEFT have already been taken into account.
We again discarded the coefficients of $\JbaseE^{(d)}$ or $\JopE^{(d)}$ respectively, assuming no contact terms among the local fields.
From this we can infer
\begin{subequations}\label{eq:coeff2Jordan}
\begin{align}
[d^{\Jbase}]_{\ctdiv} &= d^{\Jop}(T^{\Jop})^{-1} + c^{\base} T^{\base\Jbase} + c^{\baseE} T^{\baseE\Jbase}, \label{eq:Jcoeff2Jordan}\\[3pt]
c^{\base}  &= c^{\op}(T^{\op})^{-1}, \\[3pt]
c^{\baseE} &= c^{\opE} - c^{\op}(T^\op)^{-1}T^{\op\opE}. \label{eq:EOMcoeff2Jordan}
\end{align}
\end{subequations}
The main lesson to learn from \eqs{eq:coeff2Jordan} is that $c^{\baseE}$ contributes to $[d^{\Jbase}]_{\ctdiv}$ and even if we initially found $c^{\opE}=0$, we could still end up with a nonzero $c^{\baseE}$ due to the renormalisation of the on-shell operator basis of the SymEFT action.
As expected, $c^{\base}$ remains unchanged when compared to the case of spectral quantities.

This has important consequences for the matching.
Firstly, while the on-shell basis of an on-shell $\ord(a)$ improved action indeed starts at $\ord(a^2)$, there may still be EOM-vanishing operators present at $\ord(a)$.
As pointed out earlier, those $\ord(a)$ terms will give rise to contact terms with any local field present, while keeping spectral quantities unaffected.
Secondly, the presence of any contact terms may give rise to non-zero tree-level matching coefficients for the local fields even if one does not find any corrections in the classical-$a$ expansion, for example for strictly local bilinears that reside only on a particular lattice site.

Eventually, we want to set $c^{\baseE}\equiv 0$ to eliminate any contact terms with \eom{} operators.
Working with an off-shell matching procedure, this can be achieved by adjusting the matching of the renormalised fundamental fields on the lattice, i.e., the gauge field, \mbox{(anti-)}quark field as well as the renormalised couplings, to those of the continuum theory\footnote{In the literature this step is commonly referred to as a field-redefinition, but a change of matching condition makes this freedom more apparent than a substitution in the path integral.}
\begin{align}
\bar{\Psi}_\mathrm{latt}(x)&\matched\bar{\Psi}_\mathrm{cont}(x)\big\{1+\sum_na^n\cev{f}{}^{(n)}(x,am_\mathrm{cont})\big\},\nonumber\\
\Psi_\mathrm{latt}(x)&\matched\big\{1+\sum_na^n\vec{f}{}^{(n)}(x,am_\mathrm{cont})\big\}\Psi_\mathrm{cont}(x),\nonumber\\
A_{\mu,\mathrm{latt}}(x)&\matched\big\{1+\sum_na^nG^{(n)}(x,am_\mathrm{cont})\big\}_{\mu\nu}A_{\nu,\mathrm{cont}}(x),\nonumber\\
m_\mathrm{latt}&\matched \big\{1+\sum_na^nb_{m}^{(n)}(am_\mathrm{cont})\big\}m_\mathrm{cont},\nonumber\\
g_\mathrm{latt}&\matched \big\{1+\sum_na^nb_{g}^{(n)}(am_\mathrm{cont})\big\}g_\mathrm{cont}
\label{eq:sketchedRedefinition}\,.
\end{align}
The subleading terms can be chosen arbitrarily with the sole constraints, that the transformation properties of the SymEFT must be kept intact and of course appropriate canonical mass-dimensions have to be chosen.
This allows, e.g., O(4) symmetry-breaking terms to be present.
We introduced here $\matched$ to indicate how one would choose the matching conditions at the level of the \emph{renormalised} fundamental fields, masses, and coupling.
Through proper use of this freedom, we can easily set in our initial example any (tree-level) matching coefficient of \eom{} operators at $\ord(a)$ in the \emph{off-shell matched} SymEFT action to zero.
As expected, this does not impact the on-shell basis of the SymEFT action at $\ord(a^2)$ because any such contributions cancel out.
In general, beyond leading order in the lattice spacing the situation becomes more difficult owing to contact terms among operators of the SymEFT action and the quadratic pieces of the change of matching conditions.
The latter will become more clear in \eq{eq:O(a)EOMsubleading}.
Therefore, at leading order in the lattice spacing the matching coefficients of the on-shell operator basis for the SymEFT action remain unchanged as do all consequences derived for spectral quantities in~\cite{Husung:2019ytz,Husung:2021mfl,Husung:2022kvi}.
Beyond tree-level and in particular for gluonic observables the off-shell matching procedure described here may need to be revisited to understand the role of gauge-fixing.

Prior to any change of matching condition the tree-level values of the three matching coefficients $d^\Jop$, $c^\op$ and $c^{\opE}$ can be obtained from the classical-$a$ expansion of the lattice action and the discretised local field.
Unfortunately, any change of matching condition for fundamental fields forming the composite local field, here quarks and anti-quarks, will already affect the matching coefficients of the operator basis $\Jbase^{(d)}$ at the first non-trivial order in the lattice spacing, i.e., in the example of remnant EOM-vanishing operators $\ord(a)$.
At tree-level and leading order in the lattice spacing, the necessary shift of (non-Jordan normal form) matching coefficients $d^{\Jop}$ can be inferred from
\begin{align}
a^d\Delta d_i^{\Jop}\Jop_i^{(d)}(x) &= a^d \left\{\Jop[\bar{\Psi}\cev{f}{}^{(d)},\Psi,A]+\Jop[\bar{\Psi},\vec{f}{}^{(d)}\Psi,A]+\Jop[\bar{\Psi},\Psi,G^{(d)}A]\right\}(x).\label{eq:dJshift}
\end{align}
In the absence of contact terms for the leading order $\Jop^{(d)}$, this can be generalised to subleading orders via an iterative strategy that keeps also track of the non-linear pieces of the change of matching condition.
Thus we can impose this strategy to eliminate remnant \eom{} operators that survived explicit Symanzik on-shell improvement of the action as well as to set all matching coefficients from \eq{eq:EOMcoeff2Jordan} (at tree-level) $c^{\baseE}\equiv 0$.
In particular, this allows to set the last term contributing to $d^{\Jbase}$ in \eq{eq:Jcoeff2Jordan} to zero, while the initial coefficients for the local fields $d^{\Jop}$ get shifted $d^{\Jop}\rightarrow d^{\Jop}+\Delta d^{\Jop}$, to account for this change in the matching condition.

Since we are working only with mass-dimension~3 fermion-bilinears up to $\ord(a^2)$ lattice artifacts, any change of matching condition to cancel operators vanishing by the gluonic EOM will play no role in \eq{eq:dJshift}.
Moreover, at 1-loop only operators that contain a single fermionic EOM will affect the on-shell basis of the local fermion bilinears, while higher powers of the EOM will only affect off-shell matching of the \eom{} operator basis of the local field.
From the list of operators in \eqs{eq:EOMaAction} and \eqref{eq:EOMa2Action} only $\opE[i\geq 2]^{(1)}$ and $\opE[i\in\{3,6,7,9,10\}]^{(2)}$ remain.

To cancel all \eom{} operators at $\ord(a)$ via a proper choice of the matching, we find
\begin{align}
(\Delta d^{\Jop})_i^{(1)}(\Jop_i^{kl})^{(1)} \EOM -\left[(c^{\baseE})^{(1)}_2\frac{m_{k+l}}{2}+(c^{\baseE})^{(1)}_3\tr(M)\right]\Jop^{kl}\,.\label{eq:O(a)EOM}
\end{align}
If we are interested in an on-shell $\ord(a)$ improved setup, i.e.,
\begin{equation}
(\Delta d^{\Jop})_i^{(1)}\equiv -(d^{\Jop})_i^{(1)}\,\wedge\,(c_j^{\base})^{(1)}\equiv 0\,\wedge (c^{\baseE})_k^{(1)}\equiv(c^{\opE})_k^{(1)},
\end{equation}
we also need to keep track of the impact on the $\ord(a^2)$ terms from the quadratic piece of the previous change of matching condition 
\begin{align}
(\Delta d^{\Jop})_i^{(1+1)}(\Jop_i^{kl})^{(2)} &\EOM -\frac{1}{16}\left\{\left[(c^{\opE})^{(1)}_2m_{k-l}\right]^2+12\left[(c^{\opE})^{(1)}_2\frac{m_{k+l}}{2}+(c^{\opE})^{(1)}_3\tr(M)\right]^2\right\}\Jop^{kl},\nonumber\\
(\Delta c^{\opE})_i^{(1+1)}(\opE)_i^{(2)}&\EOM -\frac{3}{4}\bar\Phi\left[(c^{\opE})_2^{(1)}M+(c^{\opE})_3^{(1)}\tr(M)\right]^2\Dslash\Phi+\ldots\,.\label{eq:O(a)EOMsubleading}
\end{align}
The ellipsis contains all other \eom{} operators that are irrelevant \emph{here} for contact terms at 1-loop order due to containing the gluonic EOM or higher powers of the fermionic EOM.
Continuing at $\ord(a^2)$ we then get
\begin{align}
(\Delta d^{\Jop})_i^{(2)}(\Jop_i^{kl})^{(2)} &\EOM (\Delta d^{\Jop})_i^{(1+1)}(\Jop_i^{kl})^{(2)}-\frac{(c^{\baseE})_3^{(2)}}{4}i\bar{q}_k\{\Gamma,\sigma_{\alpha\beta}\}F_{\alpha\beta}q_l\nonumber\\
&\hphantom{\EOM}-\Bigg\{\sum_{i=3,6}\Big[(c^{\baseE})_i^{(2)}+(\Delta c^{\opE})_i^{(1+1)}\Big]\frac{m_{k+l}^2+m_{k-l}^2}{4}\nonumber\\
&\hphantom{\EOM-\Bigg\{}+(c^{\baseE})_7^{(2)}\tr(M^2)+\Big[(c^{\baseE})_9^{(2)}+(\Delta c^{\opE})_9^{(1+1)}\Big]\tr(M)\frac{m_{k+l}}{2}\nonumber\\
&\hphantom{\EOM-\Bigg\{}+\Big[(c^{\baseE})_{10}^{(2)}+(\Delta c^{\opE})_{10}^{(1+1)}\Big]\tr(M)^2\Bigg\}\Jop^{kl},\label{eq:dJshiftHere}
\end{align}
where we also included the case of an on-shell $\ord(a)$ improved theory.
The (non-trivial) anti-commutators relevant here are
\begin{align}
\{\gamma_5,\sigma_{\alpha\beta}\}F_{\alpha\beta}&=2\sigma_{\alpha\beta}\tilde{F}_{\alpha\beta},&
\{\gamma_\mu,\sigma_{\alpha\beta}\}F_{\alpha\beta}&=4i\gamma_5\gamma_\alpha\tilde{F}_{\alpha\mu},\nonumber\\
\{\gamma_5\gamma_\mu,\sigma_{\alpha\beta}\}F_{\alpha\beta}&=4i\gamma_\alpha\tilde{F}_{\alpha\mu},&
\{\sigma_{\mu\nu},\sigma_{\alpha\beta}\}F_{\alpha\beta}&=4\gamma_5\tilde{F}_{\mu\nu}+4F_{\mu\nu}.
\end{align}
To summarise, we eventually obtain for the matching coefficients
\begin{subequations}\label{eq:coeff2JordanNoEOM}
\begin{align}
\left.[d^{\Jbase}]_{\ctdiv}\right|_{c^{\baseE}=0} &= (d^{\Jop}+\Delta d^{\Jop})(T^{\Jop})^{-1} + c^{\base} T^{\base\Jbase}, \label{eq:Jcoeff2JordanNoEOM}\\[3pt]
c^{\base}  &= c^{\op}(T^{\op})^{-1}.
\end{align}
\end{subequations}
Any matching coefficient from \eq{eq:Jcoeff2JordanNoEOM} that vanishes to $(n_i^{\Jbase}-1)$th loop order will shift the truly-leading power in $\gbar^2(1/a)$ of the particular contribution according to
\begin{equation}
\hat{\Gamma}^{\Jop}_i=\hat{\gamma}_i^{\Jop}+n_i^{\Jbase}.
\end{equation}
Lacking any knowledge of the matching coefficients beyond tree-level, we will assume here $n_i^\Jbase=1$ if the matching coefficient vanishes at tree-level.
This is the same convention used for the on-shell basis $\base_i^{(d)}$ when we were discussing spectral quantities~\cite{Husung:2021mfl,Husung:2022kvi}.

%% file: examples.tex
\section{Some examples}\label{sec:examples}
For the examples we will focus on $\Nf=3$ Wilson quarks~\cite{Wilson:1974,Wilson:1975id}
\begin{align}
\hat{S}_\mathrm{W}&=a^4\sum_{x}
\begin{pmatrix}
\hat{\bar{\Psi}}\\
\hat{\bar{\Phi}}
\end{pmatrix}
\Big[\frac{\gamma_\mu}{2}\left\{\hat{\nabla}_\mu+\hat{\nabla}_\mu^*\right\}\left\{1-\hat{c}_\text{cube}(g_0)a^2\hat\Delta_\mu\right\}-\frac{ar}{2}\hat\Delta_\mu+\hat{M}_0\left\{1+\hat{b}_m(g_0)a\hat{M}_0\right\}\nonumber\\*
&\hphantom{=}+\hat{c}_\text{SW}(g_0)\frac{ia}{4}\sigma_{\mu\nu}\hat{F}_{\mu\nu}\Big]
\begin{pmatrix}
\hat{\Psi}\\
\hat{\Phi}
\end{pmatrix}(x)
\end{align}
in both sea and valence with identical choice for the Wilson term ($r=1$) combined with the L\"uscher-Weisz gauge action~\cite{Luscher:1984xn}.
Here $\hat\nabla_\mu$ and $\hat\nabla_\mu^*$ are the covariant forward and backward lattice derivatives respectively and 
\begin{equation}
\hat\Delta_\mu=\hat\nabla_\mu\hat\nabla_\mu^*\,.
\end{equation}
$\hat{c}_\text{SW}(g_0)$ is the improvement coefficient for the Sheikholeslami-Wohlert term~\cite{Sheikholeslami:1985ij}, $\hat{b}_m(g_0)$ is the improvement coefficient for the quark-mass, and $\hat{c}_\text{cube}(g_0)$ is the only additional improvement coefficient needed to achieve on-shell $\ord(a^2)$ improvement at tree-level, see also~\cite{Alford:1996nx}.
All these coefficients are here assumed to be chosen identically in the sea and valence, which minimises the operator basis further by imposing $\text{SU}(\Nf+\Nb|\Nb)_\mathrm{V}$ graded flavour symmetries in the massless limit.
Of course any other choice of discretisation compatible with the initially imposed flavour symmetries is accessible as well.
For convenience we will denote lattice quantities as their closely related continuum counterparts with a hat added.
This choice yields for the tree-level matching coefficients of the SymEFT action, see also~\cite{Husung:2019ytz,Husung:2021mfl},
\begin{align*}
\textbf{\boldmath $\ord(a)$:} &&& c^\op=(\hat{c}_\text{SW}(0)-1,0,\hat{b}_m(0)-1/2,0,\ldots,0),\quad c^{\opE}=(-1/2,1,0),\\*
\textbf{\boldmath $\ord(a^2)$:} &&& c^\op=(0,0,1/6-\hat{c}_\text{cube}(0),0,\ldots,0),\quad c^{\opE}=(0,\ldots,0),
\end{align*}
where the ordering is the one from \app{sec:onshellActionBasis} and \ref{sec:EOMmixing}.
All tree-level matching-coefficients can be easily obtained via the naive classical-$a$ expansion.
We stop here at the leading order depending on whether $\hat{c}_\text{SW}$ (and $\hat{b}_m$) has been set~\cite{Luscher:1996sc} to achieve on-shell $\ord(a)$ improvement of the action.
We \emph{assume} here that on-shell $\ord(a)$ improvement of the lattice action does not introduce additional $\ord(a^2)$ effects at tree-level beyond contact terms.
This assumption is reasonable because commonly used functional forms for the improvement coefficients incorporate the appropriate tree-level matching coefficients, see e.g. the non-perturbative determination of $\hat{c}_\text{SW}(g_0)$ in \cite{Luscher:1996ug}.

\paragraph{Example 1:} Trivially-flavoured vector 2-point function\\
To not obfuscate the discussion of lattice artifacts with the need for renormalisation let us first discuss the conserved vector current as proposed in~\cite{Frezzotti:2001ea,Heitger:2020zaq}
\begin{align}
\hat{\text{V}}_\mu^{kl}&=\bar{q}_k\gamma_\mu q_l -\frac{a}{4}\bar{q}_k\left\{\hat\nabla_\mu+\hat\nabla_\mu^*-(\hat\nabla_\mu+\hat\nabla_\mu^*)^\dagger\right\}q_l+\frac{a^2}{4}\bar{q}_k\gamma_\mu\left\{\hat\Delta_\mu+\hat\Delta_\mu^\dagger\right\}q_l\nonumber\\*
&-\hat{c}_\text{V}(g_0)\frac{a}{2}(\hat{\partial}_\nu+\hat{\partial}_\nu^*)i\bar{q}_k\sigma_{\nu\mu}q_l,\label{eq:conservedVector}
\end{align}
where $\hat{c}_\text{V}(g_0)$ is the only improvement coefficient needed at $\ord(a)$.
Its tree-level value is $\hat{c}_\text{V}(0)=1/2$, for a fully non-perturbative determination see~\cite{Heitger:2020zaq}.
$\hat{\partial}_\nu$ and $\hat{\partial}_\nu^*$ are the lattice forward and backward derivatives.
The absence of any massive $\ord(a)$ improvement-terms has been worked out in \cite{Frezzotti:2001ea} and will serve here as a non-trivial check for our tree-level matching procedure, i.e., we need to find vanishing $\ord(am)$ contributions in the SymEFT description of the conserved vector.

The first step is to work out the classical-$a$ expansion of $\hat{\text{V}}_\mu$, which we can directly read off from \eq{eq:conservedVector} using
\begin{equation}
\bar{q}_k\big\{\cev{D}_\mu-D_\mu\big\}q_l=\partial_\nu \text{T}_{\nu\mu}^{kl}+m_{k+l}\text{V}_\mu^{kl}+\Veom_1^{(1)},
\end{equation}
where $\Veom_1^{(1)}$ is a \eom{} operator with vector quantum numbers that can be found in \app{sec:EOMmixing}.
This yields for the tree-level matching coefficients
\begin{align*}
\textbf{\boldmath $\ord(a)$:} &&& d^{\hat{\text{V}}}=(1/2-\hat{c}_\text{V}(0),1,0),\\*
\textbf{\boldmath $\ord(a^2)$:} &&& d^{\hat{\text{V}}}=(1/4,0,\ldots,0).
\end{align*}
We are here free to ignore any \eom{} operators as those contribute only to subleading powers in the lattice spacing via contact terms with operators from the SymEFT action.

The quantity of interest here is the vector 2-point function
\begin{equation}
\hat{G}(x_0)=-a^3\sum_{\vecx}\big\langle \hat{\jmath}_i(x_0,\vecx)\hat{\jmath}_i(0)\big\rangle,
\end{equation}
where $x_0>0$ in the continuum limit and
\begin{equation}
3\hat{\jmath}_i(x)=2\hat{\text{V}}_i^{uu}(x)-\hat{\text{V}}_i^{dd}(x).
\end{equation}
This particular choice is inspired by the window-quantities computed for the hadronic vacuum polarization contribution to the muon anomalous magnetic moment, see e.g.~\cite{RBC:2018dos,Aubin:2022hgm,Ce:2022kxy,FermilabLattice:2022izv,RBC:2023pvn}.
After diagonalising the operator basis following the procedure described in \sects{sec:OpsRenorm} and \ref{sec:TLmatching} we then find at $\ord(a)$
\begin{align}
\frac{\hat{G}(x_0)}{\displaystyle\lim_{a\searrow 0}\hat{G}(x_0)} = 1+a\Big\{&\left[2b_0\gbar^2(1/a)\right]^{4/27}\left[4 (1-\hat{c}_\text{SW}(0)) + (1/2-\hat{c}_\text{V}(0))\right]G_{1;\text{RGI}}^{(1)}(x_0)\nonumber\\
+&\ord\left(\left[2b_0\gbar^2(1/a)\right]^{31/27}\right)\Big\}\nonumber\\
-a\Big\{&\left[2b_0\gbar^2(1/a)\right]^{-5/9}(\hat{c}_\text{SW}(0)-1)\delta G_{1;\text{RGI}}^{(1)}(x_0)\nonumber\\
+&\left[2b_0\gbar^2(1/a)\right]^{2/27}(\hat{c}_\text{SW}(0)-1)\delta G_{2;\text{RGI}}^{(1)}(x_0)\nonumber\\
+&\ord\left(\left[2b_0\gbar^2(1/a)\right]^{4/9}\right)\Big\}\nonumber\\
+\ord(&a^2)\,.
\end{align}
Notice that in the second curly brackets, only terms that are less suppressed in powers of the running coupling compared to the first subleading correction have been written out explicitly.
There are two shorthands distinguishing corrections to the local field itself and insertions of an operator of the SymEFT action (with contact divergences subtracted), i.e.,
\begin{subequations}\label{eq:correctionTerms}
\begin{align}
G_{n;\text{RGI}}^{(d)}&=2\frac{\int\rmd^3x\,\langle(\jmath_i)_{n;\text{RGI}}^{(d)}(x_0,\vecx)\jmath_i(0)\rangle}{\int\rmd^3x\,\langle\jmath_i(x_0,\vecx)\jmath_i(0)\rangle}\,,\\
\delta G_{n;\text{RGI}}^{(d)}&=\frac{\int\rmd^3x\,\rmd^4z\,\langle\jmath_i(x_0,\vecx)\jmath_i(0)\base_{n;\text{RGI}}^{(d)}(z)\rangle_\ctdiv}{\int\rmd^3x\,\langle\jmath_i(x_0,\vecx)\jmath_i(0)\rangle}\,.
\end{align}
\end{subequations}
The overall normalisation of the various contributions depends on the conventions used for defining the diagonalised bases.
We are primarily interested in identifying contributions present at tree-level.
For the same reason we write the overall factors in a way that easily allows to identify the required improvement coefficient(s) to cancel each contribution.
As expected, we find no need for a massive improvement term for the point-split vector at $\ord(a)$ and tree-level.
Once all other improvement coefficients have been set to their appropriate values one achieves tree-level $\ord(a)$ improvement and the loop-suppressed contributions take over.
Assuming now full on-shell $\ord(a)$ improvement one may repeat the same analysis at $\ord(a^2)$.\footnote{Due to a significantly larger basis of operators at $\ord(a^2)$, bringing the 1-loop anomalous dimension matrix into Jordan normal form can no longer be achieved symbolically exact in a reasonable amount of time (in \texttt{Mathematica}). 
Instead one may choose to perform the last step with finite accuracy.
For details we refer to the supplemental material.}
Here it is important to first keep track of the quadratic piece of the change of matching condition \eq{eq:O(a)EOMsubleading} when absorbing any $\ord(a)$ \eom{} operators.
The matching coefficients at $\ord(a^2)$ depend on the particular choices made for the discretisation of the tensor current used in \eq{eq:conservedVector}.
For any such choice the analysis is very similar to the second example.

It should be clear that $\hat{G}(x_0)$ requires use of some scale-setting parameter on the lattice to make it dimensionless prior to taking the continuum limit.
Each choice of scale-setting parameter will of course introduce its own set of lattice artifacts alongside those discussed here.
Similarly fixing the quark masses will introduce lattice artifacts as well.

\paragraph{Example 2:} Pseudo-scalar decay constant\\
A commonly used discretisation of the axial-vector~\cite{Luscher:1996sc} generalised to possibly non-degenerate valence quarks reads
\begin{equation}
\hat{\mathrm{A}}_\mu^{k\neq l}=\left\{1+\hat{b}_\text{A}(g_0)a\frac{\hat{m}_{k+l}}{2}+\hat{\bar{b}}_\text{A}(g_0)a\tr(\hat{M})\right\}\left\{\bar{q}_k\gamma_\mu\gamma_5q_l+\hat{c}_\text{A}(g_0)\frac{a}{2}(\hat{\partial}_\mu+\hat{\partial}_\mu^*)\bar{q}_k\gamma_5q_l\right\}.
\end{equation}
Here $\hat{b}_\text{A}(g_0)$, $\hat{\bar{b}}_\text{A}(g_0)$ and $\hat{c}_\text{A}(g_0)$ are improvement coefficients, and $\hat{m}_{k+l}$ and $\tr(\hat{M})$ denote the subtracted quark masses analogous to our continuum conventions.
Expanding the axial-vector naively in lattice spacing then yields
\begin{align*}
\textbf{\boldmath $\ord(a)$:} &&& d^{\hat{\text{A}}}=(\hat{c}_\text{A}(0),\hat{b}_\text{A}(0),\hat{\bar{b}}_\text{A}(0)),\\*
\textbf{\boldmath $\ord(a^2)$:} &&& d^{\hat{\text{A}}}=(0,\ldots,0,\hat{c}_\text{A}(0)\hat{b}_\text{A}(0),0,\hat{c}_\text{A}(0)\hat{\bar{b}}_\text{A}(0),0,0,0).
\end{align*}
Unfortunately, Wilson QCD breaks chiral symmetry explicitly and we have to renormalise the axial-vector in the lattice theory.
The renormalised pseudo-scalar decay constant then is defined (up to the corresponding pseudo-scalar meson mass as an overall factor) in the usual way
\begin{equation}
\widehat{m_\text{X} f_\text{X}}=Z_{\hat{\text{A}}}(g_0)\langle 0|\hat{A}_0^{k\neq l}|\text{X}(\vecn)\rangle,\quad \text{X}\in\{\pi,\mathrm{K}\}.
\end{equation}
To avoid the discussion of renormalisation once again we simply discuss the ratio of two pseudo-scalar decay constants, here for the Pion and Kaon,
\begin{equation}
\hat{R}(a)=\frac{\widehat{m_\mathrm{K} f_\mathrm{K}}}{\widehat{m_\pi f_\pi}}.
\end{equation}
It should be obvious that $Z_{\hat{\text{A}}}(g_0)$ cancels out in this ratio.
Assuming now full Symanzik $\ord(a)$ improvement of both the lattice action and the axial-vector combined, we find for this ratio
\begin{align}
\frac{\hat{R}(a)}{\lim\limits_{a'\searrow 0}\hat{R}(a')}=1&+a^2\sum_j[2b_0\gbar^2(1/a)]^{\hat{\Gamma}_j^{\text{A}}}\left(f_{j;\mathrm{K}}^{(2)}-f_{j;\pi}^{(2)}\right)\nonumber\\*
&-a^2\sum_j[2b_0\gbar^2(1/a)]^{\hat{\Gamma}_j^{\base}}\left(\delta f_{j;\mathrm{K}}^{(2)}-\delta f_{j;\pi}^{(2)}\right)\nonumber\\*
&+\ord\left(a^2[2b_0\gbar^2(1/a)]^{\hat{\Gamma}_\text{sub}^{\text{A},\base}},a^3\right),
\end{align}
where the RGI shorthands used here are analogous to the ones in \eqs{eq:correctionTerms} but all leading order matching coefficients have been absorbed and identical powers in $\gbar^2(1/a)$ have been summed up.
The distinct leading powers $\hat{\Gamma}_j^{\text{A}}$ and $\hat{\Gamma}_j^{\base}$ can be read off from \tab{tab:powersAxial} up to the first subleading power.
It is interesting to notice that simply setting $\hat{c}_\text{cube}(0)=1/6$ would shift all these powers but $\hat{\Gamma}_5^{\text{A}}$ by at least +1 as they can all be traced back to this one operator in the SymEFT action either through renormalisation of the basis for the action or contact terms.
Thus tree-level Symanzik $\ord(a^2)$ improvement of the action would make $\hat{\Gamma}_5^{\text{A}}$ and $\hat{\Gamma}_\text{sub}^{\base}$ the asymptotically leading contributions here.
Dividing out the overall ratio of the two pseudo-scalar meson masses would again introduce another set of lattice artifacts but restricted to contributions from the SymEFT action due to being spectral quantities.
Clearly, this would only modify the factors accompanying $[2b_0\gbar^2(1/a)]^{\hat{\Gamma}_j^{\base}}$.

\begin{table}
\caption{Distinct leading powers in $\gbar^2(1/a)$ below the first subleading power modifying classical $a^2$ behaviour as $a\searrow 0$ for the $\ord(a)$ improved local axial-vector $\ord(a)$ improved $\Nf=3$ Wilson QCD.
Discretisation of the sea and valence sector are assumed to be identical.
\underline{Underlined} numbers belong to massive contributions.
\dotuline{Underdotted} numbers correspond to massive contributions, that vanish in the mass-degenerate case.
$\hat{\Gamma}_\text{sub}^{\text{A},\base}$ denotes the first subleading power in $\gbar^2(1/a)$ due to 1-loop corrections.}\label{tab:powersAxial}\centering
\begin{tabular}{l|ccccc|r}
$j$                         & 1 & 2 & 3 & 4 & 5 & $\hat{\Gamma}_\text{sub}^{\text{A},\base}$\\[1pt]\hline\hline
$\hat{\Gamma}_j^{\text{A}}$ & $0$ & $0.395$ & $\dotuline{0.593}$ & $0.617$ & $\underline{0.889}$ & $1$\\
$\hat{\Gamma}_j^{\base}$    & $\underline{-0.111}$ & $0.247$ & $\underline{0.519}$ & $0.668$ & $0.760$ & $0.795$
\end{tabular}
\end{table}

%% file: concl.tex
\section{Discussion}\label{sec:discussion}
One lesson to be learned (again) is that on-shell Symanzik-improvement of the lattice action in general also benefits local fields due to the absence of contact-terms apart from remnant EOM-vanishing operators.
Combined with tree-level Symanzik-improvement of the local fields, this forces all matching coefficients to vanish at tree-level.\footnote{For any perturbative determination of improvement coefficients involving off-shell contributions, any remnant EOM operators must be absorbed first!}
Furthermore, continuum extrapolations in small volume are free of $\ord(a)$ lattice artifacts in the chiral limit due to the mass-dimension~4 operators found here having opposite chirality --- the $\ord(a)$ correction to any non-trivial continuum matrix element simply vanishes by chirality arguments.
This also implies that any $\ord(a)$ effects in the near-massless finite-volume theory should be suppressed in powers of the renormalised quark masses.
Here, the vanishing of $\ord(a)$ corrections does not imply absence of the corresponding operators in our minimal basis, but is due to automatic $\ord(a)$ improvement very similar to, e.g., maximally twisted Wilson quarks~\cite{Frezzotti:2003ni,Aoki:2006gh,Sint:2007ug}.
If there is no $\ord(a)$ improvement they will have an impact on the $\ord(a^2)$ lattice artifacts by the interplay with $\ord(a)$ terms from other local fields (or the SymEFT action).
In case of an infinite (sufficiently large) volume, dynamical chiral-symmetry breaking invalidates any chirality arguments mentioned earlier.

The powers reported for $\hat{\gamma}^J$ (and previously $\hat{\gamma}^\base$~\cite{Husung:2021mfl,Husung:2022kvi}) are universal for lattice actions that realise the symmetries assumed here but further suppression due to vanishing matching coefficients may arise depending on the particular formulation.
We denote this by $\hat{\Gamma}^J$ and $\hat{\Gamma}^\base$ respectively.
Using different discretisations for the local fields simultaneously for a continuum extrapolation, gives only a fairly limited handle on the powers $\hat{\Gamma}^J$, while contributions from the SymEFT action with powers~$\hat{\Gamma}^{\base}$ remain unchanged.
While being much more expensive, having two (or more) different lattice actions at hand allows for some check of universality and a better control on the quality of a combined continuum extrapolation.

This work serves primarily to highlight the correct Symanzik treatment of local fields with all its pitfalls.
To some, the role of EOM-vanishing operators in the SymEFT action may come as a surprise as one usually drops those from the beginning.
That this is in general not correct in the presence of local fields should be clear at this point and the proper treatment has been outlined in the previous sections.
This has already been pointed out in the past~\cite{Capitani:1999ay,Capitani:2000xi} for GW quarks and is relevant for on-shell local fields despite the papers discussing Symanzik off-shell improvement.
It should be clear that any Symanzik-improvement of local fields must take such effects into account.
Here, the final change of basis followed by a change of the matching condition keeps track of any impact of such operators.

Aside from the explicit numbers given here, all cases covered in this paper are accessible via the attached \texttt{Mathematica} notebooks.
Unfortunately, the full strategy outlined here is very tedious and must be repeated for \emph{any} local field of interest.
Only the mixing of the operators in the SymEFT action can be reused.\footnote{Since we rely on the proper bookkeeping of \eom{} operators one must work in the same gauge, here implemented via the background field method~\cite{tHooft:1975uxh,Abbott:1980hw,Abbott:1981ke,Luscher:1995vs}.}
For each local field the derivation of a minimal basis and the renormalisation thereof, including the renormalisation of contact-terms, must be done repeatedly.
In particular, deriving the minimal basis for the local field is very tedious unless some kind of automation can be devised.
What may be beneficial is to use a slightly over-complete basis instead and postpone the full reduction of the basis after the renormalisation.
Using an over-complete basis, a minimal-subtraction scheme (and likely any scheme) will highlight any redundant terms by the occurrence of redundancies in the choices for the counter-terms hinting at linear dependencies among the operators.
Although helpful, such a procedure will not help to distinguish between operators that should be absorbed into the set of \eom{} operators and those to be kept as minimal on-shell basis.

Following \eq{eq:gammaJ}, the anomalous dimension of the continuum local field may provide guidance on which local fields should be prioritised when looking for distinctly negative powers in $\gbar^2(1/a)$.
If the anomalous dimension of the continuum local field itself is very positive it will shift the overall power $\hat{\Gamma}^J$ of the higher-dimensional basis towards negative values and therefore make distinctly negative powers more likely.
One such example to keep in mind are the 4-quark operators discussed in~\cite{Ciuchini:1997bw} relevant for $\Delta F=2$ effective Hamiltonians, where at least one 1-loop anomalous dimension is enlarged.
Of course this should only be taken as a crude guideline as one can never be certain what 1-loop anomalous dimensions one will find for the higher-dimensional basis.

\section{Limitations}\label{sec:limitations}
Throughout this work we assumed use of a lattice quark action that has (at least) $\text{SU}(\Nf)_\mathrm{V}\times\text{SU}(\Nb|\Nb)_\mathrm{V}$ graded flavour symmetry in the massless limit and preserves discrete rotations, charge, parity, and time reversal.
Our results are therefore applicable to both Wilson quarks~\cite{Wilson:1974,Wilson:1975id} and GW quarks~\cite{Ginsparg:1981bj}, where the latter impose an even stronger constraint due to exact lattice chiral symmetry in the massless limit.
This stronger constraint can be enforced by dropping the appropriate higher-dimensional local fields from our (massless) basis.
For lattice quark actions violating any of these symmetry constraints, one has to revisit the derivation of the minimal operator bases in \sect{sec:minBasis} as well as the minimal operator basis relevant for the SymEFT action.

Less obvious subtleties arise because of the restriction to the on-shell basis for the local fields.
For any renormalisation scheme that relies on \emph{off-shell renormalisation conditions} like, e.g., RI/(S)MOM schemes~\cite{Martinelli:1994ty,Sturm:2009kb} one can no longer treat the EOM-vanishing basis of the local fields as irrelevant.
Although this problem has been identified before~\cite{Bhattacharya:2000ch,Constantinou:2009tr,Gockeler:2010yr}, overall awareness in the literature seems to be faint.
Keep in mind that the powers computed here \emph{hold only for on-shell matrix elements} and are therefore incomplete to describe lattice artifacts of any $Z$-factors that have been determined via an off-shell renormalisation condition.
The latter may also have gauge-choice-dependent lattice artifacts~\cite{Giusti:2002rn}.
The proper treatment of gauge-choice-dependent lattice artifacts in a SymEFT is beyond the scope of this paper and will be very complicated due to losing gauge-symmetry as a constraint on the minimal operator basis.
In contrast, for example the Schr\"odinger functional~\cite{Luscher:1992an,Sint:1993un} allows one to define an on-shell non-perturbative renormalisation scheme~\cite{Jansen:1995ck} and will at most require additional operators for the SymEFT action on the time-boundaries to be taken into account, see also~\cite{Luscher:1992an,Luscher:1996sc,Husung:2019ytz}.

Moreover, for \emph{integrated correlation functions}, e.g. moments~\cite{Bochkarev:1995ai,HPQCD:2008kxl} or the hadronic contributions to muon $g-2$, namely the hadronic vacuum-polarisation, and the hadronic light-by-light contribution, the operator bases derived here will only be relevant as a subset.
The presence of contact terms of the local fields in the lattice theory gives rise to divergences on the lattice that need to be renormalised.
On the SymEFT side the EOM-vanishing basis of the local fields become relevant and even more powers in $\gbar^2(1/a)$ will arise from the contact term contributions.
For example in the case of integrated 2-point functions, contact terms in the lattice theory will give rise to contributions from the flavour-singlet scalar just by taking quantum numbers and the canonical mass-dimensions at the contact-interaction into account.
This restriction may be relaxed for so-called \emph{window-quantities} at intermediate or long distances due to  sufficient suppression of contact terms rendering them less potent to cause problems.
Recently, those window-quantities have gained quite some interest as a benchmark for the hadronic vacuum-polarisation contribution to muon $g-2$, see e.g.~\cite{RBC:2018dos,Aubin:2022hgm,Ce:2022kxy,FermilabLattice:2022izv,RBC:2023pvn}.

\section{Conclusion}\label{sec:conclusion}
We have computed the additional asymptotically leading powers in $\gbar^2(1/a)$ that one encounters for mass-dimension~3 fermion bilinears, except for the scalar with trivial flavour quantum numbers.

Again, no seriously negative powers are found for the cases most commonly used in the literature assuming (at least tree-level) $\ord(a)$ improvement of the local fields and non-trivial flavour quantum numbers.
For an unimproved valence Wilson action, the tensor has a negative power at $\ord(a)$ with $\min_i(\hat{\gamma}^\mathrm{T})_i^{(1)}\sim -0.14$ due to a contact term, which is however very close to the classical case. 
Meanwhile, finding $\min_i(\hat{\gamma}^\mathrm{A})_i^{(1)}\sim -0.4$ at $\ord(a)$ for the axial-vector is less of an issue since this contribution has by construction a vanishing tree-level matching coefficient for the commonly used strictly-local discretisation of the axial-vector --- this highlights the importance of taking potential suppression from tree-level matching into account.

Otherwise, for trivial flavour quantum numbers the pseudo-scalar at $\ord(a)$ and the axial-vector at $\ord(a^2)$ give rise to negative powers in $\gbar^2(1/a)$ with $\min_i(\hat{\gamma}^{\mathrm{P}})_i^{(1)}\sim -0.6$ and $\min_i(\hat{\gamma}^{\mathrm{A}})_i^{(2)}= -1$.
All these cases can and should be remedied by at least tree-level Symanzik improvement.
The powers found here combined with those from the SymEFT action can now in principle be used for ans\"atze of the leading asymptotic lattice spacing dependence in continuum extrapolations of decay constants, form factors etc., that is for matrix elements of the local bilinears discussed here.
As before, one caveat remains owing to the presence of $\ord(a^{\nmin+1})$ contributions as well as the question whether leading-order perturbative predictions are sufficient in the range of lattice spacings available.

Let us stress again the importance of any remnant \eom{} operators present at $\ord(a)$ after on-shell Symanzik improvement of the action.
Those will give rise to $\ord(a)$ terms of the local field even in the absence of explicit $\ord(a)$ terms in the classical-$a$ expansion of the local field.
It should be clear that this is true independent of the choice of matching condition.
For the off-shell matching strategy discussed in this paper such effects are accounted for by the appropriate changes of matching in the fundamental fields~\eq{eq:O(a)EOM}.
For an on-shell strategy this is automatically taken care of.

All results presented here should generalise to more diverse choices in the valence sector.
As long as each set of flavours is compatible with at least the lattice symmetries of Wilson quarks this simply enlarges the operator bases further but should not lead to any new powers $\hat{\Gamma}^{J,\base}$.
Due to an increased degeneracy of the 1-loop anomalous dimensions being found, more $\log(2b_0\gbar^2(1/a))$ factors modifying the leading powers in $\gbar^2(1/a)$ may arise.

Central computational steps described here have been implemented in \texttt{FORM}~\cite{Vermaseren:2000nd} scripts, Python scripts and \texttt{Mathematica} notebooks including some automation via a \texttt{Makefile}, all of which is publicly available.\footnote{\url{https://github.com/nikolai-husung/Symanzik-QCD-workflow}}%
Adaptation to other choices of local fields should be straight forward, but some changes might be needed, e.g., if the number of free spacetime indices increases compared to the tensor current.

\paragraph{Supplementary material.}
Alongside this manuscript a \texttt{Mathematica} notebook is supplied to obtain the leading powers in $\gbar^2(1/a)$ for our particular choice of lattice action with $\Nf$ flavours in the sea and $\Nb$ valence quarks.
Optionally, the tree-level matching coefficients of the Jordan normal form of the basis for the SymEFT can be obtained by providing the coefficients of the \emph{classical-$a$ expansion} for both the lattice action and the local field including EOM vanishing operators.

\paragraph{Acknowledgements.}
I am grateful to Rainer Sommer, Chris Sachrajda, and Jonathan Flynn for discussions on the automatic $\ord(a)$ improvement of the massless theory in finite volume and thank Jonathan Flynn, Andreas J\"uttner, Rainer Sommer, and Gregorio Herdoíza for comments and discussions on various stages of the manuscript.
The author acknowledges funding by the STFC consolidated grant ST/T000775/1 as well as support of the projects PID2021-127526NB-I00, funded by MCIN/AEI/10.13039/501100011033 and by FEDER EU, IFT Centro de Excelencia Severo Ochoa No CEX2020-001007-S, funded by MCIN/AEI/10.13039/501100011033, H2020-MSCAITN-2018-813942 (EuroPLEx), under grant agreement No. 813942, and the EU Horizon 2020 research and innovation programme, STRONG-2020 project, under grant agreement No. 824093.
The Feynman diagrams used in this paper have been generated with help of the \texttt{LaTeX} package \texttt{TikZ-Feynman}~\cite{Ellis:2016jkw}.

%% file: appendix.tex
\appendix

\section{Listing of the minimal on-shell basis for the SymEFT action}\label{sec:onshellActionBasis}
The full on-shell basis of the SymEFT action up to $\ord(a^2)$ has been derived before and we are only reusing our previous choices~\cite{Husung:2022kvi}.
The on-shell basis at mass-dimension~5 reads
\begin{align}
\op_1^{(1)}&=\frac{i}{4}\bar{\chi}\sigma_{\mu\nu}F_{\mu\nu}\chi, &
\op^{(1)}_2&=\tr(M)\frac{1}{\bare{g}^2}\tr(F_{\mu\nu}F_{\mu\nu}),&
\op^{(1)}_3&=\bar{\chi}M^2\chi,\nonumber\\
\op^{(1)}_4&=\tr(M)\bar{\chi}M\chi,&
\op^{(1)}_5&=\tr(M^2)\bar{\chi}\chi,\vphantom{\frac{1}{g_0^2}} &
\op^{(1)}_6&=\tr(M)^2\bar{\chi}\chi.
\end{align}
Here, $\tr(M^n)$ always denotes a trace over the sea-quark mass-matrix.
For mass-dimension~6 we use
\begin{align}
\op^{(2)}_{1}&=\frac{1}{\bare{g}^2}\tr([D_\mu, F_{\nu\rho}]\,[D_\mu, F_{\nu\rho}])\,, &
\op^{(2)}_{2}&=\frac{1}{\bare{g}^2}\sum\limits_{\mu}\tr([D_\mu, F_{\mu\nu}]\,[D_\mu, F_{\mu\nu}])\,,\nonumber\\
\op^{(2)}_3&=\sum_\mu\bar\chi\gamma_\mu D_\mu^3\chi,&
\op^{(2)}_4&=\bare{g}^2(\bar\chi\gamma_\mu\chi)^2,\nonumber\\
\op^{(2)}_5&=\bare{g}^2(\bar\chi\gamma_\mu\gamma_5\chi)^2,&
\op^{(2)}_6&=\bare{g}^2(\bar\chi\gamma_\mu T^a\chi)^2,\nonumber\\
\op^{(2)}_7&=\bare{g}^2(\bar\chi\gamma_\mu\gamma_5T^a\chi)^2, &
\op^{(2)}_8&=\bare{g}^2(\bar\chi\chi)^2,\nonumber\\
\op^{(2)}_9&=\bare{g}^2(\bar\chi\gamma_5\chi)^2,&
\op^{(2)}_{10}&=\bare{g}^2(\bar\chi\sigma_{\mu\nu}\chi)^2,\nonumber\\
\op^{(2)}_{11}&=\bare{g}^2(\bar\chi T^a\chi)^2,&
\op^{(2)}_{12}&=\bare{g}^2(\bar\chi\gamma_5T^a\chi)^2,\nonumber\\
\op^{(2)}_{13}&=\bare{g}^2(\bar\chi\sigma_{\mu\nu}T^a\chi)^2,&
\op^{(2)}_{14}&=\frac{i}{4}\bar{\chi}M\sigma_{\mu\nu}F_{\mu\nu}\chi,\nonumber\\
\op^{(2)}_{15}&=\tr(M^2)\frac{1}{\bare{g}^2}\tr(F_{\mu\nu}F_{\mu\nu}),&
\op^{(2)}_{16}&=\bar{\chi}M^3\chi,\nonumber\\
\op^{(2)}_{17}&=\tr(M^2)\bar{\chi}M\chi,&
\op^{(2)}_{18}&=\frac{i\tr(M)}{4}\bar{\chi}\sigma_{\mu\nu}F_{\mu\nu}\chi,\nonumber\\
\op^{(2)}_{19}&=\tr(M)^2\frac{1}{\bare{g}^2}\tr(F_{\mu\nu}F_{\mu\nu}),&
\op^{(2)}_{20}&=\tr(M)\bar{\chi}M^2\chi,\nonumber\\
\op^{(2)}_{21}&=\tr(M)^2\bar{\chi}M\chi,&
\op^{(2)}_{22}&=\tr(M^3)\bar{\chi}\chi,\nonumber\\
\op^{(2)}_{23}&=\tr(M^2)\tr(M)\bar{\chi}\chi,&
\op^{(2)}_{24}&=\tr(M)^3\bar{\chi}\chi,
\end{align}
where we introduced $\chi=\Psi,\Phi$ as a flavour vector in the valence or sea sector.
Also the mixed variants of 4-quark operators are needed
\begin{equation}
\op^{(2)}_{j\in\{4,\ldots,13\};\text{sea-val}}=g_0^2\bar\Psi\Gamma_j\Psi\bar\Phi\Gamma_j\Phi\,.
\end{equation}
The latter operators play an important role for contact terms with the SymEFT action as they connect the valence with the sea sector.
For convenience, we choose here a slightly modified normalisation of the massive bases by dropping overall factors of $1/\Nf$.
Notice that all $\op_i^{(1)}$ as well as $\op^{(2)}_{i\not\in\{1,2,3,4,5,6,7,14,15,16,17\}}$ are not invariant under the spurion symmetry transformation 
\begin{align}
M & \rightarrow R M L^\dagger,\quad \chi_\mathrm{R}=\frac{1+\gamma_5}{2}\chi \rightarrow R \chi_\mathrm{R}, \quad \chi_\mathrm{L}=\frac{1-\gamma_5}{2}\chi \rightarrow L \chi_\mathrm{L},\quad \bar\chi_\mathrm{R} \rightarrow \bar\chi_\mathrm{R} R^\dagger,\nonumber\\
&\bar\chi_\mathrm{L} \rightarrow \bar\chi_\mathrm{L} L^\dagger,\quad R \in \SU(\Nf)_\mathrm{R},\quad L \in \SU(\Nf)_\mathrm{L},\label{eq:chiralSpurion}
\end{align}
where the subscripts L and R refer to left-handed and right-handed fermions respectively.
This spurion symmetry limits the allowed operator mixing severely.
Also, for any lattice action preserving chiral symmetry in the massless limit, operators incompatible with the spurion symmetry are therefore forbidden.
Consequently, those operators will not contribute for lattice actions with exact lattice chiral symmetry~\cite{Luscher:1998pqa}.

\section{Listing of \eom{} operators}\label{sec:EOMmixing}
To work out the (tree-level) matching coefficients for the leading order lattice artifacts of the local fields, we need some insight into the mixing of \eom{} operators under renormalisation.
Once we know how the \eom{} operators mix into the on-shell basis of the SymEFT action as well as how they mix within \eom{} operators, we can work out the matching coefficients $c^{\baseE}$ in~\eq{eq:EOMcoeff2Jordan}.
With this knowledge, we can eventually adjust the matching conditions according to \eq{eq:sketchedRedefinition} to get rid of any \eom{} operator present in the minimal basis of the SymEFT action.

Before we can do this, we give here the minimal basis of \eom{} operators
\begin{align}
\opE[1]^{(1)} &= \bar\chi \Dslash^2 \chi, &
\opE[2]^{(1)} &= \bar\chi M \Dslash \chi, &
\opE[3]^{(1)} &= \tr(M) \bar\chi \Dslash \chi, \label{eq:EOMaAction}\\
\opE[1]^{(2)} &= \frac{1}{g_0^2}\tr(D_\mu F_{\mu\rho} D_\nu F_{\nu\rho})+\mathrlap{\frac{1}{2}\bar\Psi \gamma_\mu D_\nu F_{\nu\mu}\Psi+\frac{1}{2}\bar\Phi \gamma_\mu D_\nu F_{\nu\mu}\Phi,} & && \nonumber\\
\opE[2]^{(2)} &= \bar\chi \gamma_\mu T^a\chi\mathrlap{\left\{D_\nu F_{\nu\mu}^a-g_0^2\bar\Psi\gamma_\mu T^a\Psi-g_0^2\bar\Phi\gamma_\mu T^a\Phi\right\},} & && 
\opE[3]^{(2)} &= \frac{1}{2}\bar\chi\big\{D^2\Dslash-\cev{\Dslash}\cev{D}^2\big\}\chi, \nonumber\\
\opE[4]^{(2)} &= \bar\chi \Dslash^3\chi, &
\opE[5]^{(2)} &= \bar\chi M \Dslash^2 \chi, &
\opE[6]^{(2)} &= \bar\chi M^2 \Dslash \chi,  \nonumber\\
\opE[7]^{(2)} &= \tr(M^2) \bar\chi \Dslash \chi, &
\opE[8]^{(2)} &= \tr(M) \bar\chi \Dslash^2 \chi, &
\opE[9]^{(2)} &= \tr(M) \bar\chi M \Dslash \chi,  \nonumber\\
\opE[10]^{(2)} &= \tr(M)^2\bar\chi \Dslash \chi.\label{eq:EOMa2Action}
\end{align}
Again, $\chi=\Psi,\Phi$ denotes a flavour-vector in the sea or valence sector in case different discretisations are chosen for both sectors.
We use here and in the following the sloppy shorthands $\Dslash q_l=(\gamma_\kappa D_\kappa + m_l)q_l$ and $\bar{q}_l\cev{\Dslash}=\bar{q}_l(\gamma_\kappa \cev{D}_\kappa - m_l)$.
Although the operators $\opE[i]^{(1)}$ and $\opE[i\geq 8]^{(2)}$ break the spurion symmetry from \eq{eq:chiralSpurion}, they are not forbidden in the case of GW quarks due to the fact that there are \emph{massive} operators present at $\ord(a)$ in the naive $a$-expansion that violate this spurion symmetry, but vanish by EOMs.
As worked out in section~\ref{sec:TLmatching}, those operators should be absorbed by a change of matching condition.
Otherwise, those operators will become relevant even for the GW action in the presence of local fields via contact interactions.
This detail clearly points to the possibility of having $\ord(a)$ terms due to contact-terms of \emph{massive} \eom{} operators of mass-dimension 5 with the local fields even when using GW quarks and ``naively'' improved local fields.
This has been discussed before~\cite{Capitani:1999ay,Capitani:2000xi} in the context of perturbative Green's functions.\footnote{Beware that the ``improved'' operator introduced there already carries the corrections from the external quark fields.}
Keep in mind that doing a field-redefinition does not eliminate those $\ord(a)$ terms, but shifts their origin from having contact terms to terms present in the minimal basis of the local field to begin with.

For completeness we also list here the minimal EOM-vanishing bases (indicated by the subscript $\mathcal{E}$) for the various local fields, which were needed due to the off-shell renormalisation strategy
\begingroup\allowdisplaybreaks
\begin{subequations}\label{eq:Jeom}
\begin{align}
\Seom^{(1)}_1&=\bar{q}_k(\cev{\Dslash}-\Dslash)q_l,&\\*
\Seom^{(2)}_1&=\bar{q}_k(\cev{\Dslash^2}+\Dslash^2)q_l,&
\Seom^{(2)}_2&=\bar{q}_k\cev{\Dslash}\Dslash q_l,\nonumber\\*
\Seom^{(2)}_3&=m_{k+l}\Seom^{(1)}_1,&
\Seom^{(2)}_4&=m_{k-l}\bar{q}_k(\cev{\Dslash}+\Dslash)q_l,\\
\PSeom^{(1)}_1&=\bar{q}_k(\cev{\Dslash}\gamma_5-\gamma_5\Dslash)q_l,\\*
\PSeom^{(2)}_1&=\bar{q}_k(\cev{\Dslash^2}\gamma_5+\gamma_5\Dslash^2)q_l,&
\PSeom^{(2)}_2&=\bar{q}_k\cev{\Dslash}\gamma_5\Dslash q_l,\nonumber\\*
\PSeom^{(2)}_3&=m_{k+l}\PSeom^{(1)}_1,&
\PSeom^{(2)}_4&=m_{k-l}\bar{q}_k(\cev{\Dslash}\gamma_5+\gamma_5\Dslash)q_l,\\*
\Veom^{(1)}_1&=\bar{q}_k(\cev{\Dslash}\gamma_\mu-\gamma_\mu\Dslash)q_l,&\\*
\Veom^{(2)}_1&=\bar{q}_k(\cev{\Dslash^2}\gamma_\mu+\gamma_\mu\Dslash^2)q_l,&
\Veom^{(2)}_2&=\bar{q}_k\cev{\Dslash}\gamma_\mu\Dslash q_l,\nonumber\\*
\Veom^{(2)}_3&=m_{k+l}\Veom^{(1)}_1,&
\Veom^{(2)}_4&=\bar{q}_k(\cev{\Dslash}\cev{D}_\mu+D_\mu\Dslash)q_l,\nonumber\\*
\Veom^{(2)}_5&=\partial_\mu\big\{\bar{q}_k(\cev{\Dslash}+\Dslash)q_l\big\},&
\Veom^{(2)}_6&=m_{k-l}\bar{q}_k(\cev{\Dslash}\gamma_\mu+\gamma_\mu\Dslash)q_l,\\
\Aeom^{(1)}_1&=\bar{q}_k(\cev{\Dslash}\gamma_5\gamma_\mu-\gamma_5\gamma_\mu\Dslash)q_l,&\\*
\Aeom^{(2)}_1&=\bar{q}_k(\cev{\Dslash^2}\gamma_5\gamma_\mu+\gamma_5\gamma_\mu\Dslash^2)q_l,&
\Aeom^{(2)}_2&=\bar{q}_k\cev{\Dslash}\gamma_5\gamma_\mu\Dslash q_l,\nonumber\\*
\Aeom^{(2)}_3&=m_{k+l}\Aeom^{(1)}_1,&
\Aeom^{(2)}_4&=\bar{q}_k(\cev{\Dslash}\cev{D}_\mu\gamma_5+\gamma_5D_\mu\Dslash)q_l\nonumber\\*
\Aeom^{(2)}_5&=\partial_\mu\PSeom^{(1)}_1,&
\Aeom^{(2)}_6&=m_{k-l}\bar{q}_k(\cev{\Dslash}\gamma_5\gamma_\mu+\gamma_5\gamma_\mu\Dslash)q_l,\nonumber\\*
\Aeom^{(2)}_7&=\delta_{kl}\mathrlap{\left(\frac{2}{\bare{g}^2}\tr(D_\rho F_{\rho\lambda}\tilde{F}_{\mu\lambda})+\bar{\Psi}\gamma_\rho\tilde{F}_{\mu\rho}\Psi+\bar{\Phi}\gamma_\rho\tilde{F}_{\mu\rho}\Phi\right),}\label{eq:axialEOM}\\
\TTeom^{(1)}_1&=i\bar{q}_k(\cev{\Dslash}\sigma_{\mu\nu}-\sigma_{\mu\nu}\Dslash)q_l,\\*
\TTeom^{(2)}_1&=i\bar{q}_k(\cev{\Dslash^2}\sigma_{\mu\nu}+\sigma_{\mu\nu}\Dslash^2)q_l,&
\TTeom^{(2)}_2&=i\bar{q}_k\cev{\Dslash}\sigma_{\mu\nu}\Dslash q_l,\nonumber\\*
\TTeom^{(2)}_3&=\partial_\mu \Veom[\nu]^{(1)}_1-\partial_\nu \Veom^{(1)}_1,&
\TTeom^{(2)}_4&=\bar{q}_k(\cev{\Dslash}\cev{D}_\mu\gamma_\nu+\gamma_\nu D_\mu\Dslash-(\mu\leftrightarrow\nu))q_l,\nonumber\\*
\TTeom^{(2)}_5&=m_{k+l}\TTeom^{(1)}_1,&
\TTeom^{(2)}_6&=im_{k-l}\bar{q}_k(\cev{\Dslash}\sigma_{\mu\nu}+\sigma_{\mu\nu}\Dslash)q_l,\nonumber\\
\TTeom^{(2)}_7&=\mathrlap{\varepsilon_{\mu\nu\rho\sigma}\partial_\rho(\bar{q}_k\{\cev{\Dslash}\gamma_5\gamma_\sigma+\gamma_5\gamma_\sigma\Dslash\})q_l.}
\end{align}
\end{subequations}
\endgroup
The operator $\Aeom^{(2)}_{7}$ must be included for completeness and induces mixing of the sea and valence singlets into the higher-dimensional axial-vector with trivial flavour quantum numbers.
As before we omitted those singlets of the sea and valence sector that can be obtained easily by summing over the flavours.\footnote{Singlets in the valence sector should here be thought of as having the full valence quark and ghost flavour-vector~$\Phi$.}

\section{1-loop block matrices for contact terms at mass-dimension~5}\label{sec:blockMatricesContact}
The block matrices of the 1-loop anomalous-dimension corresponding to contact terms at mass-dimension~5 are
\begin{subequations}
\begin{align}
&\frac{(4\pi)^2\Nc}{\Nc^2-1}\left[\gamma_0^{\op,\mathrm{S}^{k\neq l}}\right]^{(2)}=\left(\arraycolsep=2pt
\begin{array}{cccc}
 \frac{12 \Nc^2}{1-\Nc^2} & 4 & -4 & -24 \\[6pt]
 \frac{3 \Nc^2}{1-\Nc^2} & \frac{4}{3} & -\frac{4}{3} & -8 \\[6pt]
 \frac{2 \left(\Nc^2-3\right)}{\Nc^2-1} & \frac{1}{3} & -\frac{13}{6} & -\frac{20}{3} \\[6pt]
 0 & \frac{8 \Nc}{\Nc^2-1} & \frac{8 \Nc}{1-\Nc^2} & \frac{48 \Nc}{1-\Nc^2} \\[6pt]
 0 & \frac{8 \Nc}{1-\Nc^2} & 0 & \frac{48 \Nc}{\Nc^2-1} \\[6pt]
 0 & -4 & 4 & 24 \\[6pt]
 0 & 4 & 0 & -24 \\[6pt]
 \frac{8 \Nc}{1-\Nc^2} & \frac{2 \Nc}{\Nc^2-1} & \frac{\Nc}{\Nc^2-1} & \frac{12 \Nc}{1-\Nc^2} \\[6pt]
 \frac{8 \Nc}{1-\Nc^2} & \frac{2 \Nc}{\Nc^2-1} & \frac{3 \Nc}{1-\Nc^2} & \frac{12 \Nc}{1-\Nc^2} \\[6pt]
 \frac{32 \Nc}{\Nc^2-1} & \frac{24 \Nc}{\Nc^2-1} & \frac{12 \Nc}{1-\Nc^2} & \frac{144 \Nc}{1-\Nc^2} \\[6pt]
 \frac{4}{1-\Nc^2} & -1 & -\frac{1}{2} & 6 \\[6pt]
 \frac{4}{1-\Nc^2} & -1 & \frac{3}{2} & 6 \\[6pt]
 \frac{16}{\Nc^2-1} & -12 & 6 & 72 \\[6pt]
 0 & 0 & \frac{3}{8} & \frac{15}{2}
\end{array}
\right),\nonumber\\
&\frac{(4\pi)^2\Nc}{\Nc^2-1}\left[\gamma_0^{\op,\mathrm{P}^{kl}}\right]^{(2)}=\left(\arraycolsep=2pt
\begin{array}{ccccc}
 \frac{12 \Nc^2}{1-\Nc^2} & 4 & -6 & 0 & -16 \\[6pt]
 \frac{3 \Nc^2}{1-\Nc^2} & \frac{4}{3} & -2 & 0 & -\frac{16}{3} \\[6pt]
 \frac{2 \left(\Nc^2-3\right)}{\Nc^2-1} & \frac{1}{3} & -\frac{11}{4} & 0 & -\frac{13}{3} \\[6pt]
\big[&\ldots& W^{kl}+\delta_{kl}W_\mathrm{val}^{k=l}&\ldots&\big]  \\[6pt]
 0 & 0 & \frac{15}{8} & \frac{2 \Nc\delta_{kl}}{\Nc^2-1} & \frac{3}{2} \\[6pt]
\big[&\ldots& \frac{\delta_{kl}}{2}W_\mathrm{sea}^{k=l}&\ldots&\big]
\end{array}
\right),\nonumber\\
&\frac{(4\pi)^2\Nc}{\Nc^2-1}\left[\gamma_0^{\op,\mathrm{V}^{kl}}\right]^{(2)}=\left(\arraycolsep=2pt\begin{array}{cccccccc}
 0 & 0 & 0 & \frac{4}{3} & -2 & \frac{8}{3} & 0 & 0 \\[6pt]
 \frac{11}{20} & \frac{11}{60} & -\frac{11}{60} & \frac{19}{40} & -\frac{57}{80} & \frac{103}{120} & -\frac{11}{60} & -\frac{11}{20} \\[6pt]
 -\frac{8}{5} & -\frac{13}{15} & -\frac{17}{15} & \frac{3}{10} & -\frac{9}{20} & \frac{31}{30} & -\frac{17}{15} & \frac{3}{5} \\[6pt]
\big[&\ldots&& \multicolumn{2}{c}{X^{kl}+\delta_{kl}X_\mathrm{val}^{k=l}}&&\ldots&\big]  \\[6pt]
 0 & 0 & 0 & 0 & \frac{3}{8} & 0 & 2 & \frac{3}{2} \\[6pt]
\big[&\ldots&& \multicolumn{2}{c}{\frac{\delta_{kl}}{2}X_\mathrm{sea}^{k=l}}&&\ldots&\big]
\end{array}
\right),\nonumber\\
&\frac{(4\pi)^2\Nc}{\Nc^2-1}\left[\gamma_0^{\op,\mathrm{A}^{kl}}\right]^{(2)}=\nonumber\\*
&\left(\arraycolsep=2pt\begin{array}{cccccccccc}
 0 & 0 & 0 & 0 & 0 & \frac{4}{3} & 0 & 0 & \frac{16}{3} & -8 \\[6pt]
 \frac{11}{20} & \frac{11}{60} & -\frac{11}{120} & 0 & -\frac{11}{60} & \frac{19}{40} & -\frac{11}{240} & 0 & \frac{103}{60} & -\frac{57}{20} \\[6pt]
 -\frac{8}{5} & -\frac{13}{15} & \frac{30\Nc\Sigma-17+17\Nc^2}{30-30\Nc^2} & \frac{4 \Nc\delta_{kl}}{1-\Nc^2} & -\frac{17}{15} & \frac{3}{10} & \frac{43}{60} & \frac{\Nc\delta_{kl}}{2 \left(\Nc^2-1\right)} & \frac{31}{15} & -\frac{9}{5} \\[6pt]
\big[&&\ldots&& \multicolumn{2}{c}{Y^{kl}+\delta_{kl}Y_\mathrm{val}^{k=l}}&&\ldots&&\big]  \\[6pt]
 0 & 0 & 1 & 0 & 0 & 0 & -\frac{5}{8} & 0 & 0 & \frac{3}{2} \\[6pt]
\big[&&\ldots&& \multicolumn{2}{c}{\frac{\delta_{kl}}{2}Y_\mathrm{sea}^{k=l}}&&\ldots&&\big]
\end{array}
\right),\nonumber\\
&\frac{(4\pi)^2\Nc}{\Nc^2-1}\left[\gamma_0^{\op,\mathrm{T}^{kl}}\right]^{(2)}=\left(\arraycolsep=2pt\begin{array}{cccccccccc}
 0 & \frac{6 \Nc^2+8}{1-\Nc^2} & \frac{6 \Nc^2+8}{1-\Nc^2} & 0 & 0 & 0 & 0 & -2 & 0 & 0 \\[6pt]
 0 & \frac{9 \Nc^2+16}{6-6 \Nc^2} & \frac{9 \Nc^2+16}{6-6 \Nc^2} & 0 & 0 & 0 & 0 & -\frac{2}{3} & 0 & 0 \\[6pt]
 0 & \frac{3 \Nc^2+11}{3-3 \Nc^2} & \frac{3 \Nc^2+11}{3-3 \Nc^2} & 0 & 0 & 0 & \frac{1}{4} & -\frac{2}{3} & 0 & 1 \\[6pt]
\big[&&\ldots&& \multicolumn{2}{c}{Z^{kl}+\delta_{kl}Z_\mathrm{val}^{k=l}}&&\ldots&&\big]  \\[6pt]
 0 & 0 & 0 & 0 & 0 & 0 & -\frac{1}{8} & \frac{1}{2} & 0 & -\frac{1}{2} \\[6pt]
\big[&&\ldots&& \multicolumn{2}{c}{\frac{\delta_{kl}}{2}Z_\mathrm{sea}^{k=l}}&&\ldots&&\big]
\end{array}
\right),\\
&\frac{(4\pi)^2\Nc}{\Nc^2-1}\left[\gamma_0^{\opE,\mathrm{S}^{k\neq l}}\right]^{(2)}=\left(\arraycolsep=2pt
\begin{array}{cccc}
 \frac{4 \left(\Nc^2-3\right)}{\Nc^2-1} & 0 & -\frac{3}{4} & -3
\end{array}
\right),\nonumber\\
&\frac{(4\pi)^2\Nc}{\Nc^2-1}\left[\gamma_0^{\opE,\mathrm{P}^{kl}}\right]^{(2)}=\left(\arraycolsep=2pt
\begin{array}{ccccc}
 \frac{4 \left(\Nc^2-3\right)}{\Nc^2-1} & 0 & -\frac{3}{4} & \frac{4 \Nc\delta_{kl}}{\Nc^2-1} & -3 
\end{array}
\right),\nonumber\\
&\frac{(4\pi)^2\Nc}{\Nc^2-1}\left[\gamma_0^{\opE,\mathrm{V}^{kl}}\right]^{(2)}=\left(\arraycolsep=2pt
\begin{array}{cccccccc}
 0 & -\frac{2}{3} & 0 & 0 & -\frac{1}{12} & \frac{1}{3} & \frac{2}{3} & 1
\end{array}
\right),\nonumber\\
&\frac{(4\pi)^2\Nc}{\Nc^2-1}\left[\gamma_0^{\opE,\mathrm{A}^{kl}}\right]^{(2)}=\left(\arraycolsep=2pt\begin{array}{cccccccccc}
 0 & \frac{4 \Nc\Sigma-2+2\Nc^2}{3-3 \Nc^2} & \frac{1}{3} & 0 & 0 & 0 & -\frac{1}{12} & 0 & \frac{2}{3} & -\frac{1}{3}
\end{array}
\right),\nonumber\\
&\frac{(4\pi)^2\Nc}{\Nc^2-1}\left[\gamma_0^{\opE,\mathrm{T}^{kl}}\right]^{(2)}=\left(\arraycolsep=2pt\begin{array}{cccccccccc}
 0 & \frac{2 \left(\Nc^2+3\right)}{1-\Nc^2} & \frac{2 \left(\Nc^2+3\right)}{1-\Nc^2} & 0 & 0 & 0 & \frac{1}{4} & 0 & 0 & 1
\end{array}
\right).
\end{align}
\end{subequations}
We dropped again explicitly massive operators of the SymEFT action, whose contact terms can be inferred from their lower-dimensional counterparts without the explicit masses.
The operators $\op$ are either gluonic or involve valence quarks.
The ordering is as in \app{sec:onshellActionBasis} and \ref{sec:EOMmixing}, with the restriction at $\ord(a^2)$ of having only $\op_{j<15}^{(2)}\cup \op_{j\in\{4,...,13\};\text{sea-val}}^{(2)}$ and $\opE[3]^{(2)}$.
The mixing of the 4-quark operators with valence-valence pairs and valence-sea pairs has been denoted using the following shorthands
\begin{align}
W^{kl}&=\left(\arraycolsep=2pt\begin{array}{ccccc}
 0 & \frac{8 \Nc}{\Nc^2-1} & \frac{12 \Nc}{1-\Nc^2} & 0 & \frac{32 \Nc}{1-\Nc^2} \\[6pt]
 0 & \frac{8 \Nc}{1-\Nc^2} & \frac{12 \Nc}{\Nc^2-1} & 0 & 0 \\[6pt]
 0 & -4 & 6 & 0 & 16 \\[6pt]
 0 & 4 & -6 & 0 & 0 \\[6pt]
 \frac{8 \Nc}{\Nc^2-1} & \frac{2 \Nc}{1-\Nc^2} & \frac{3 \Nc}{\Nc^2-1} & 0 & \frac{4 \Nc}{1-\Nc^2} \\[6pt]
 \frac{8 \Nc}{\Nc^2-1} & \frac{2 \Nc}{1-\Nc^2} & \frac{3 \Nc}{\Nc^2-1} & 0 & \frac{12 \Nc}{\Nc^2-1} \\[6pt]
 \frac{32 \Nc}{1-\Nc^2} & \frac{24 \Nc}{1-\Nc^2} & \frac{36 \Nc}{\Nc^2-1} & 0 & \frac{48 \Nc}{\Nc^2-1} \\[6pt]
 \frac{4}{\Nc^2-1} & 1 & -\frac{3}{2} & 0 & 2 \\[6pt]
 \frac{4}{\Nc^2-1} & 1 & -\frac{3}{2} & 0 & -6 \\[6pt]
 \frac{16}{1-\Nc^2} & 12 & -18 & 0 & -24
\end{array}
\right),\quad 
W_x^{k=l}=\left(\arraycolsep=2pt\begin{array}{ccccc}
 0 & 0 & 0 & 0 & 0 \\[6pt]
 0 & 0 & 0 & 0 & \frac{8 \Nc^2m_{k+l}^2\Sigma_x\partial_{m^2}}{1-\Nc^2} \\[6pt]
 0 & 0 & 0 & 0 & 0 \\[6pt]
 0 & 0 & 0 & 0 & 0 \\[6pt]
 0 & 0 & 0 & 0 & 0 \\[6pt]
 0 & \frac{8 \Nc^2\Sigma_x}{\Nc^2-1} & 0 & 0 & \frac{4 \Nc^2m_{k+l}^2\Sigma_x\partial_{m^2}}{1-\Nc^2} \\[6pt]
 0 & 0 & 0 & 0 & 0 \\[6pt]
 0 & 0 & 0 & 0 & 0 \\[6pt]
 0 & 0 & 0 & 0 & 0 \\[6pt]
 \frac{32 \Nc\Sigma_x}{\Nc^2-1} & 0 & 0 & 0 & 0
\end{array}
\right),\nonumber\\
X^{kl}&=\left(\arraycolsep=2pt\begin{array}{cccccccc}
 0 & 0 & 0 & \frac{8 \Nc}{3 \left(\Nc^2-1\right)} & -\frac{4 \Nc}{\Nc^2-1} & \frac{16 \Nc}{3 \left(\Nc^2-1\right)} & 0 & 0 \\[6pt]
 0 & 0 & 0 & \frac{8 \Nc}{3 \left(\Nc^2-1\right)} & \frac{4 \Nc}{1-\Nc^2} & \frac{32 \Nc}{3-3 \Nc^2} & 0 & 0 \\[6pt]
 0 & 0 & 0 & -\frac{4}{3} & 2 & -\frac{8}{3} & 0 & 0 \\[6pt]
 0 & 0 & 0 & -\frac{4}{3} & 2 & \frac{16}{3} & 0 & 0 \\[6pt]
 0 & 0 & 0 & \frac{4 \Nc}{3-3 \Nc^2} & \frac{2 \Nc}{\Nc^2-1} & \frac{10 \Nc}{3 \left(\Nc^2-1\right)} & \frac{4 \Nc}{\Nc^2-1} & 0 \\[6pt]
 0 & 0 & 0 & \frac{4 \Nc}{3 \left(\Nc^2-1\right)} & \frac{2 \Nc}{1-\Nc^2} & \frac{2 \Nc}{3 \left(\Nc^2-1\right)} & \frac{4 \Nc}{\Nc^2-1} & 0 \\[6pt]
 0 & 0 & 0 & 0 & 0 & \frac{24 \Nc}{\Nc^2-1} & \frac{16 \Nc}{1-\Nc^2} & 0 \\[6pt]
 0 & 0 & 0 & \frac{2}{3} & -2 & -\frac{5}{3} & -2 & 4 \\[6pt]
 0 & 0 & 0 & -\frac{2}{3} & 1 & -\frac{1}{3} & -2 & 0 \\[6pt]
 0 & 0 & 0 & 0 & 0 & -12 & 8 & 0
\end{array}
\right),\nonumber\\
X_x^{k=l}&=\left(\arraycolsep=2pt\begin{array}{cccccccc}
 0 & 0 & 0 & \frac{16 \Nc^2\Sigma_x}{3 \left(\Nc^2-1\right)} & 0 & 0 & 0 & 0 \\[6pt]
 0 & 0 & 0 & 0 & 0 & 0 & 0 & 0 \\[6pt]
 0 & 0 & 0 & 0 & 0 & 0 & 0 & 0 \\[6pt]
 0 & 0 & 0 & 0 & 0 & 0 & 0 & 0 \\[6pt]
 0 & 0 & 0 & 0 & 0 & 0 & 0 & 0 \\[6pt]
 0 & 0 & 0 & 0 & 0 & 0 & 0 & 0 \\[6pt]
 0 & 0 & 0 & 0 & 0 & 0 & \frac{16 \Nc^2m_{k+l}\Sigma_x\partial_m}{1-\Nc^2} & 0 \\[6pt]
 0 & 0 & 0 & 0 & 0 & 0 & 0 & 0 \\[6pt]
 0 & 0 & 0 & 0 & 0 & 0 & 0 & 0 \\[6pt]
 0 & 0 & 0 & 0 & 0 & 0 & 0 & 0 
\end{array}
\right),\nonumber\\
Y^{kl}&=\left(\arraycolsep=2pt\begin{array}{cccccccccc}
 0 & 0 & 0 & 0 & 0 & \frac{8 \Nc}{3 \left(\Nc^2-1\right)} & 0 & 0 & \frac{32 \Nc}{3 \left(\Nc^2-1\right)} & \frac{16 \Nc}{1-\Nc^2} \\[6pt]
 0 & 0 & 0 & 0 & 0 & \frac{8 \Nc}{3 \left(\Nc^2-1\right)} & 0 & 0 & \frac{64 \Nc}{3-3 \Nc^2} & \frac{16 \Nc}{1-\Nc^2} \\[6pt]
 0 & 0 & 0 & 0 & 0 & -\frac{4}{3} & 0 & 0 & -\frac{16}{3} & 8 \\[6pt]
 0 & 0 & 0 & 0 & 0 & -\frac{4}{3} & 0 & 0 & \frac{32}{3} & 8 \\[6pt]
 0 & 0 & \frac{2 \Nc}{1-\Nc^2} & 0 & 0 & \frac{4 \Nc}{3 \left(\Nc^2-1\right)} & \frac{2 \Nc}{\Nc^2-1} & 0 & \frac{20 \Nc}{3-3 \Nc^2} & \frac{8 \Nc}{1-\Nc^2} \\[6pt]
 0 & 0 & \frac{2 \Nc}{1-\Nc^2} & 0 & 0 & \frac{4 \Nc}{3-3 \Nc^2} & \frac{2 \Nc}{\Nc^2-1} & 0 & \frac{4 \Nc}{3-3 \Nc^2} & \frac{8 \Nc}{\Nc^2-1} \\[6pt]
 0 & 0 & \frac{8 \Nc}{\Nc^2-1} & 0 & 0 & 0 & \frac{8 \Nc}{1-\Nc^2} & 0 & \frac{48 \Nc}{1-\Nc^2} & 0 \\[6pt]
 0 & 0 & 1 & 0 & 0 & -\frac{2}{3} & -1 & 0 & \frac{10}{3} & 4
\end{array}
\right),\nonumber\\
Y_x^{k=l}&=\left(\arraycolsep=2pt\begin{array}{cccccccccc}
 0 & 0 & 0 & 0 & 0 & 0 & 0 & 0 & 0 & 0 \\[6pt]
 0 & 0 & 0 & 0 & 0 & \frac{16 \Nc^2\Sigma_x}{3 \left(\Nc^2-1\right)} & 0 & 0 & \frac{16 \Nc^2m_{k+l}\Sigma_x\partial_m}{3-3 \Nc^2} & \frac{8 \Nc^2m_{k+l}^2\Sigma_x\partial_{m^2}}{1-\Nc^2} \\[6pt]
 0 & 0 & 0 & 0 & 0 & 0 & 0 & 0 & 0 & 0 \\[6pt]
 0 & 0 & 0 & 0 & 0 & 0 & 0 & 0 & 0 & 0 \\[6pt]
 0 & 0 & 0 & 0 & 0 & 0 & 0 & 0 & 0 & 0 \\[6pt]
 0 & 0 & 0 & 0 & 0 & 0 & 0 & 0 & \frac{8 \Nc^2m_{k+l}\Sigma_x\partial_{m}}{\Nc^2-1} & 0 \\[6pt]
 0 & 0 & 0 & 0 & 0 & 0 & 0 & 0 & 0 & 0 \\[6pt]
 0 & 0 & 0 & 0 & 0 & 0 & 0 & 0 & 0 & 0 \\[6pt]
 0 & 0 & 0 & 0 & 0 & 0 & 0 & 0 & 0 & 0 \\[6pt]
 0 & 0 & 0 & 0 & 0 & 0 & 0 & 0 & 0 & 0 
\end{array}
\right),\nonumber\\
Z^{kl}&=\left(\arraycolsep=2pt\begin{array}{cccccccccc}
 0 & \frac{16 \Nc}{\Nc^2-1} & \frac{16 \Nc}{\Nc^2-1} & 0 & 0 & 0 & 0 & \frac{4 \Nc}{1-\Nc^2} & \frac{8 \Nc}{1-\Nc^2} & 0 \\[6pt]
 0 & \frac{16 \Nc}{1-\Nc^2} & \frac{16 \Nc}{1-\Nc^2} & 0 & 0 & 0 & 0 & \frac{4 \Nc}{1-\Nc^2} & \frac{8 \Nc}{1-\Nc^2} & 0 \\[6pt]
 0 & \frac{8}{\Nc^2-1} & \frac{8}{\Nc^2-1} & 0 & 0 & 0 & 0 & 2 & 4 & 0 \\[6pt]
 0 & \frac{8}{1-\Nc^2} & \frac{8}{1-\Nc^2} & 0 & 0 & 0 & 0 & 2 & 4 & 0 \\[6pt]
 0 & \frac{4 \Nc}{\Nc^2-1} & \frac{4 \Nc}{1-\Nc^2} & 0 & \frac{2 \Nc}{3 \Nc^2-3} & \frac{4 \Nc}{3-3 \Nc^2} & \frac{\Nc}{\Nc^2-1} & \frac{2 \Nc}{1-\Nc^2} & 0 & \frac{4 \Nc}{1-\Nc^2} \\[6pt]
 0 & \frac{4 \Nc}{\Nc^2-1} & \frac{4 \Nc}{1-\Nc^2} & 0 & \frac{2 \Nc}{3\Nc^2-3} & \frac{4 \Nc}{3-3 \Nc^2} & \frac{\Nc}{\Nc^2-1} & \frac{2 \Nc}{\Nc^2-1} & \frac{4 \Nc}{1-\Nc^2} & \frac{4 \Nc}{1-\Nc^2} \\[6pt]
 0 & \frac{48 \Nc}{\Nc^2-1} & \frac{48 \Nc}{1-\Nc^2} & 0 & \frac{8 \Nc}{3-3 \Nc^2} & \frac{16 \Nc}{3 \Nc^2-3} & \frac{4 \Nc}{1-\Nc^2} & 0 & 0 & \frac{16 \Nc}{\Nc^2-1} \\[6pt]
 0 & \frac{2}{\Nc^2-1} & \frac{2}{1-\Nc^2} & 0 & -\frac{1}{3} & \frac{2}{3} & -\frac{1}{2} & 1 & -2 & 2 \\[6pt]
 0 & \frac{2}{\Nc^2-1} & \frac{2}{1-\Nc^2} & 0 & -\frac{1}{3} & \frac{2}{3} & -\frac{1}{2} & -1 & 2 & 2 \\[6pt]
 0 & \frac{24}{\Nc^2-1} & \frac{24}{1-\Nc^2} & 0 & \frac{4}{3} & -\frac{8}{3} & 2 & 0 & 0 & -8 
\end{array}
\right),\nonumber\\
Z_x^{k=l}&=\left(\arraycolsep=2pt\begin{array}{cccccccccc}
 0 & 0 & 0 & 0 & 0 & 0 & 0 & 0 & \frac{8 \Nc^2m_{k+l}\Sigma_x\partial_m}{1-\Nc^2} & 0 \\[6pt]
 0 & 0 & 0 & 0 & 0 & 0 & 0 & 0 & 0 & 0 \\[6pt]
 0 & 0 & 0 & 0 & 0 & 0 & 0 & 0 & 0 & 0 \\[6pt]
 0 & 0 & 0 & 0 & 0 & 0 & 0 & 0 & 0 & 0 \\[6pt]
 0 & 0 & 0 & 0 & 0 & 0 & 0 & 0 & 0 & 0 \\[6pt]
 0 & 0 & 0 & 0 & 0 & 0 & 0 & 0 & 0 & 0 \\[6pt]
 0 & 0 & 0 & 0 & \frac{16 \Nc^2\Sigma_x}{3-3 \Nc^2} & \frac{32 \Nc^2\Sigma_x}{3 \Nc^2-3} & 0 & 0 & 0 & \frac{8 \Nc^2m_{k+l}^2\Sigma_x\partial_{m^2}}{\Nc^2-1} \\[6pt]
 0 & \frac{8 \Nc\Sigma_x}{\Nc^2-1} & 0 & 0 & 0 & 0 & 0 & 0 & 0 & 0 \\[6pt]
 0 & 0 & \frac{8 \Nc\Sigma_x}{1-\Nc^2} & 0 & 0 & 0 & 0 & 0 & 0 & 0 \\[6pt]
 0 & 0 & 0 & 0 & 0 & 0 & 0 & 0 & 0 & 0
\end{array}
\right),
\end{align}
where $m_{k+l}^n\Sigma_x\partial_{m^n}$ implies removing the explicit quark-mass dependence from the operator before summing over all flavours $x\in\{\text{val},\text{sea}\}$ and replace it with the proper valence quark masses.
Appropriate factors of 2 compatible with our conventions for $m_{k+l}$ have been applied.
While the block matrices given here contain all the information needed, a less dense form can be found in the \texttt{Mathematica} package in the supplemental material.

\lstset{basicstyle=\ttfamily,language=Mathematica,upquote=true}
\section{How to use the supplemental material}
To ease utilisation of the supplemental material, we highlight here the general use case on the basis of the two examples in Wilson QCD in \sect{sec:examples}.
The first step always requires the setup of the supplied package
\begin{lstlisting}[caption={Setting up the package.}]
SetDirectory[NotebookDirectory[]];
<< localfieldsMixing`
\end{lstlisting}
assuming the \texttt{Mathematica} notebook to be in the same directory as the package.
The central functions to be used in the following are
\begin{lstlisting}[caption={Query function arguments.}]
?getDiagLocalContact
(*
> getDiagLocalContact[Ncvalue,Nfvalue,symm,d:1,trivialFlavour:False,numericalJordan:False,xivalue:1]
> ...
*)
?makeFieldRedef
(*
> makeFieldRedef[coeffsEOM,Nfvalue,symm,d:1,trivialFlavour:False,subleading:False]
> ...
*)
\end{lstlisting}
where the full explanation can be obtained by running the above commands.

If we were interested in the trivially-flavoured vector at $\ord(a)$ we could run
\begin{lstlisting}[caption={TL matching of the vector in unimproved Wilson QCD.},label={lst:vector}]
Ncvalue=3;
Nfvalue=3;
dvalue=1;
{gammaHat,dgammaHat,T,Tinv,bases}=getDiagLocalContact[Ncvalue, Nfvalue, "V", dvalue, True];
dVqq = {1/2 - cV, 1, 0, 0, 0, 0, 0, 0, 0, 0, 0};
cOnshell = {0, cSWsea - 1, bm - 1/2, 0, 0, 0, cSWval - 1, bm - 1/2, 0, 0, 0};
cEOM = {-1/2, 1, 0, -1/2, 1, 0};
(* Compute matching coefficients of the basis in Jordan normal form. *);
c = Transpose[Tinv] . Join[dVqq, -cOnshell, -cEOM];
{deltaJ, deltaJsub, deltaOsub} = makeFieldRedef[-c[[Range[Length[dVqq] + Length[cOnshell] + 1, Length[c]]]], Nfvalue, "V", dvalue, True];
(* Take appropriate change of matching condition into account. *)
cFinal = Transpose[Tinv] . Join[dVqq + deltaJ, -cOnshell, -cEOM - c[[Range[1 + Length[dVqq] + Length[cOnshell], Length[c]]]]];
\end{lstlisting}
The provided tree-level coefficients have to match the ordering returned in \texttt{bases}. 
Eventually, \texttt{cFinal} holds the matching coefficients, which can be traced back through the change of basis \texttt{T} to either insertions of operators from the SymEFT action or higher-dimensional local fields.
To each matching coefficient comes a power in $\gbar^2(1/a)$ stored in \texttt{gammahat} while \texttt{dgammahat} should be checked for nonzero entries hinting at the occurrence of explicit $\log(2b_0\gbar^2(1/a))$ factors at leading order.

Similarly we can run for the second example, i.e., the axial-vector
\begin{lstlisting}[caption={TL matching of the axial-vector in $\ord(a)$ improved Wilson QCD.},label={lst:axial}]
Ncvalue = 3;
Nfvalue = 3;
(* First absorb O(a) remnant EOM terms. *)
dvalue = 1;
{gammaHat, dgammaHat, T, Tinv, bases} = getDiagLocalContact[Ncvalue, Nfvalue, "A", dvalue, False, False];
dAqQ = {cA, bA, bAbar};
cOnshell = {0, cSWsea - 1, bm - 1/2, 0, 0, 0, cSWval - 1, bm - 1/2, 0, 0, 0};
cEOM = {-1/2, 1, 0, -1/2, 1, 0};
c = Transpose[Tinv] . Join[dAqQ, -cOnshell, -cEOM];
{deltaJ, deltaJsub, deltaOsub} = makeFieldRedef[-c[[Range[Length[dAqQ] + Length[cOnshell] + 1, Length[c]]]], Nfvalue, "A", dvalue, False, True];
cFinal = Transpose[Tinv] . Join[dAqQ + deltaJ, -cOnshell, -cEOM -c[[Range[1 + Length[dAqQ] + Length[cOnshell], Length[c]]]]];
(* Now continue at O(a^2) *)
dvalue = 2;
{gammaHat, dgammaHat, T, Tinv, bases} = getDiagLocalContact[Ncvalue, Nfvalue, "A", dvalue, False, True];
dAqQ = ConstantArray[0, 12] + deltaJsub;
cOnshell = ConstantArray[0, 54];
cOnshell[[5]] = cOnshell[[25]] = 1/6 - cCube;
cEOM = ConstantArray[0, 19] + deltaOsub;
c = Transpose[Tinv] . Join[dAqQ, -cOnshell, -cEOM];
{deltaJ, deltaJsub, deltaOsub} = makeFieldRedef[-c[[Range[Length[dAqQ] + Length[cOnshell] + 1, Length[c]]]], Nfvalue, "A", dvalue, False];
cFinal = Transpose[Tinv] . Join[dAqQ + deltaJ, -cOnshell, -cEOM - c[[Range[1 + Length[dAqQ] + Length[cOnshell], Length[c]]]]]/.{cA->0, cSWsea->1, cSWval->1, bAbar->0, bA->1, bm->1/2};
\end{lstlisting}
Due to the significantly enlarged operator basis, we perform the last step of the Jordan decomposition numerically, here indicated by the last argument being \texttt{True} when calling \texttt{getDiagLocalContact}.